\documentclass[aps,amsmath,twocolumn,amssymb,floatfix,showpacs,superscriptaddress,longbibliography]{revtex4-1}
\usepackage{graphicx}
\usepackage[dvipsnames]{xcolor}
\usepackage{float}
\usepackage[normalem]{ulem}
\usepackage{amsmath}
\newcommand{\stkout}[1]{\ifmmode\text{\sout{\ensuremath{#1}}}\else\sout{#1}\fi}
\usepackage{cancel,soul,ulem}
\usepackage{amsmath, bigstrut}

\usepackage[colorlinks=true,linktoc=page,linkcolor=magenta,citecolor=blue,urlcolor=purple]{hyperref}
\usepackage{ulem}

\usepackage{soul}
\usepackage{color}
\usepackage{comment}

\newcommand{\ie}{{\it i.e.,\,\,}}
\newcommand{\eg}{{\it e.g.,~}}

\newcommand\bea{\begin{eqnarray}}
\newcommand\eea{\end{eqnarray}}
\newcommand\beq{\begin{equation}}  
\newcommand\eeq{\end{equation}}

\begin{document}



\title{Thermoelectric properties of inversion symmetry broken Weyl semimetal-Weyl superconductor hybrid junctions}

\author{Ruchi Saxena}
\altaffiliation{The first three authors contributed equally to this work}
     \affiliation{Advanced Technology Institute and Department of Physics, University of Surrey, Guildford, GU2 7XH, United Kingdom}
     \affiliation{National Physical Laboratory, Hampton Road, Teddington TW11 0LW, United Kingdom}
        \author{Nirnoy Basak}
\altaffiliation{The first three authors contributed equally to this work}
     \affiliation{Harish-Chandra Research Institute, Chhatnag Road, Jhunsi, Allahabad, Uttar Pradesh 211 019, India}
     \affiliation{Homi Bhabha National Institute, Training School Complex, Anushakti Nagar, Mumbai 400094, India}
     \author{Pritam Chatterjee}
     \altaffiliation{The first three authors contributed equally to this work}
     \affiliation{Institute of Physics, Sachivalaya Marg, Bhubaneswar, Orissa 751005, India}
     \affiliation{Homi Bhabha National Institute, Training School Complex, Anushakti Nagar, Mumbai 400094, India}
     \author{Sumathi Rao}
     \email{sumathi.rao@icts.res.in}
     \affiliation{International Centre for Theoretical Sciences (ICTS-TIFR), Shivakote, Hesaraghatta Hobli, Bangalore 560089, India}
    \affiliation{Harish-Chandra Research Institute, Chhatnag Road, Jhunsi, Allahabad, Uttar Pradesh 211 019, India}
    \author{Arijit Saha}
    \email{arijit@iopb.res.in}
    \affiliation{Institute of Physics, Sachivalaya Marg, Bhubaneswar, Orissa 751005, India}
    \affiliation{Homi Bhabha National Institute, Training School Complex, Anushakti Nagar, Mumbai 400094, India}




\begin{abstract}
    \noindent
    We theoretically investigate the thermoelectric properties (electronic contribution) of a hybrid structure comprising of an inversion symmetry broken Weyl semimetal (WSM) and             
    intrinsic Weyl superconductor (WSC) with $s$-wave pairing, employing the Blonder-Tinkham-Klapwijk formulation for non-interacting electrons. Our study unfolds interesting 
    features for various relevant physical quantities such as the thermal conductance, the thermoelectric coefficient and the corresponding figure of merit. We also explore the effects of an 
    interfacial insulating (I) barrier (WSM-I-WSC set-up) on the thermoelectric response in the thin barrier limit. Further, we compute the ratio of the thermal to the electrical conductance in 
    different temperature regimes and find that the Wiedemann-Franz law is violated for small temperatures (below critical temperature $T_{c}$) near the Weyl points while it saturates to the 
    Lorentz number,  away from the Weyl points, at all temperatures irrespective of the barrier strength. We compare and contrast this behaviour with other Dirac material heterostructures 
    and provide a detailed analysis of the thermal transport. Our study can facilitate the fabrication of mesoscopic thermoelectric devices based on WSMs.
\end{abstract}

\maketitle

\section{Introduction} \label{introduction}
In recent times, Weyl semimetals (WSMs) have been subject to intense theoretical and experimental investigations as they are explicit material realizations of hitherto high energy phenomena 
such as the Adler-Bell-Jackiw anomaly~\cite{Adler,BellJackiw} and the chiral magnetic effect~\cite{PhysRevD.78.074033}.  In terms of their band structure, WSMs exhibit linearly dispersing  
excitations from non-degenerate band touching points called Weyl nodes accompanying unusual surface projections known as Fermi arcs~\cite{Armitage2018}. Weyl nodes always appear in pairs 
of opposite chirality and behave as oppositely charged monopoles of Berry flux because the electronic states around the band touching points have non-zero Berry curvature. This gives rise to 
non-trivial topology in momentum space due to which WSMs exhibit several interesting physical effects such as the quantum anomalous Hall effect, the chiral magnetic effect, negative 
magneto-resistance, etc.~\cite{Armitage2018,Hasan2015,2Hasan2015,Franz2013,Burkov2012}.  Intense theoretical and experimental research work has been carried out in both time-reversal 
and inversion symmetry broken WSMs~\cite{Armitage2018,Hasan2015,Ding2005}. 
  
 In principle, the WSM phase can be realized by breaking either the time reversal (TR) or the inversion symmetry (IS)~\cite{Armitage2018,Burkov2011, Franz2013,Claudia2017}. Breaking of the TR   
 symmetry requires a large external magnetic field, which limits the use and application of these materials. However, recently, a WSM state was confirmed in transition metal mono-phosphides or 
 mono-arsenides MX materials (M=Nb and Ta; X=P and As) with naturally broken IS due to crystal structure assymetry~\cite{Yun2019,Yu2016,Hasan2015,Yan2015,Ding2015,Claudia2015}. In these 
 materials, the Weyl nodes and surface Fermi arcs are detected by using angle-resolved photoemission spectroscopy (ARPES). 

Over the past years, superconducting hybrid structures, have attracted a great deal of 
    attention due to the dramatic boosts of thermoelectric effects in them~\cite{Chandrasekhar_2009,Machon_2014,PhysRevLett.112.057001,PhysRevLett.109.147004,PhysRevLett.110.047002,PhysRevLett.116.097001,Arijit2017,SJozef}.
In order to investigate the thermoelectric properties of a material or hybrid junction, it is 	 	
     customary to compute the thermal conductance or thermal current generated by the applied
    temperature gradient. From the application point of view, it is more desirable to investigate the  
   Seebeck coefficient, known as thermopower. 
A better way to examine the efficiency of a system as a thermoelectric is to study the thermopower as well as
a dimensionless parameter called the figure of merit ($zT$)~\cite{C1EE02497C}. 
Improving this thermoelectric $zT$ along with 
  enhanced Seebeck coefficient is one of the main challenges in material science~\cite{Snyder2008} as well as mesoscopic hybrid junctions~\cite{RevModPhys.78.217}. 
  Along this direction, heat transport has also been investigated in superconducting heterostructures of two-dimensional Dirac systems~\cite{Linder2008,Arijit2016,Razieh,Marzari2016,
  PhysRevB.81.113401,Zare2019}.

Although electronic properties of WSMs (both in bulk and hetero-junctions) have been extensively studied in recent times~\cite{Uchida2014,SRAK2016,SRAK2017,Trauzettel2018,PhysRevB.104.075420,Paramita2020}, 		far less is known about its 
thermoelectric properties as far as hybrid junctions are concerned. There have been earlier studies including disorder and interactions both near the Weyl point and with doping away from the Weyl 
point~\cite{Yun2019,Nicklas2017,Gregory2016,Gregory2014}. However, these works only investigate thermoelectric properties in the bulk material. Hetero-junctions which form the foundation of 
applications in electronics and spintronics, have not yet been explored in the context of its thermal properties involving WSMs and superconductivity. 
Motivated by this fact, in this article, we focus on this gap and explicitly study the thermal properties of a hetero-junction consisting of an inversion symmetry broken WSM (normal region) on one side  
and a bulk Weyl superconductor (WSC)~\cite{PhysRevB.86.054504,PhysRevB.86.214514,PhysRevB.92.035153} on the other side, thus tailoring a WSM-WSC junction. In particular, we obtain the 
electronic contribution to the thermal conductivity, the Seebeck coefficient and the figure of merit ($\sim$ the ratio of the Seebeck 
coefficient and the thermal conductivity) for this setup. Note that a similar analysis cannot be carried out for a TR broken WSM-WSC junction, as Andreev reflection is fully suppressed there in the 
absence of a spin active interface, due to chirality blockade~\cite{CBBeenakker}.

The remainder of this paper is organized as follows. In Sec.~\ref{ModelHamiltonian}, we describe the model Hamiltonian of IS broken WSM and the scattering matrix approach to analyse our setup. 
Sec.~\ref{thermaltransport} is devoted to the analytical formulae for computing various physical quantities that are required to assess the thermoelectric properties of the system. 
We discuss the numerical results of the WSM-WSC junction in Sec.~\ref{numerical results}.
Finally, we summarize our findings and discuss some possible outlooks in Sec.~\ref{Summary and Discussion}.

 \section{Model and Method} \label{ModelHamiltonian}
In this section we describe the model Hamiltonian of our setup and discuss the scattering matrix approach to analyse the junction problem. 
    \subsection{Model Hamiltonian}

\begin{figure}[h!]
    \begin{center}
        \includegraphics[width=3.7in,height=2.2in]{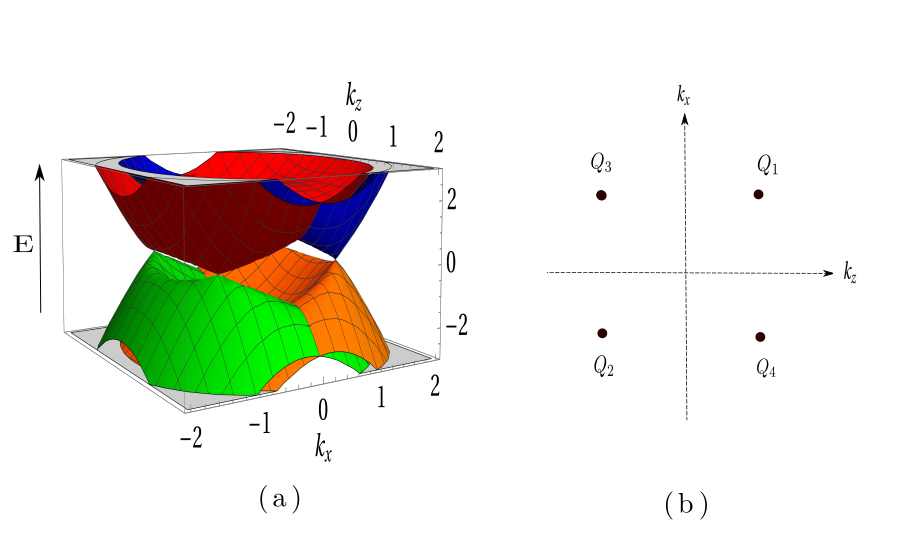}
    \end{center}
    \caption{\hspace*{-0.1cm}(a) Band structure of IS broken WSM, modelled by the low energy effective Hamiltonian (Eq.~(\ref{WeylHamil})), is shown at $k_y=0$ 
    for fixed values of $\beta=1$ and $\alpha=\sqrt{2}$. 
        (b) The positions of the Weyl nodes ($Q_{\gamma=1,2,3,4}$) are shown schematically in the $k_{x}$-$k_{z}$ plane choosing $k_{y}=0$.
        }
      \label{bandstructure}
\end{figure}

To begin with, we consider an inversion asymmetric WSM described by the Hamiltonian~\cite{Trauzettel_PRL}, $\mathcal{H}= \sum_{\bf{k}} \psi_{\bf{k}}^\dag H({\bf{k}}) \psi_{\bf{k}}$ with  
\begin{eqnarray}
H({\bf{k}})&=&k_x \sigma_x s_z + k_y \sigma_y s_0+(k_0^2-|{\bf{k}}|^2)\sigma_z s_0 \nonumber\\
&&+\beta\, \sigma_y s_y-\alpha\, k_y \sigma_x s_y\ ,
\label{WeylHamil}
\end{eqnarray}
where, ${\bf{k}}=(k_{x}, k_{y}, k_{z})$, $\psi_{{\bf{k}}}^\dag = (c_{A,\uparrow,{\bf{k}}}^\dag, c_{A,\downarrow,{\bf{k}}}^\dag, c_{B,\uparrow,{\bf{k}}}^\dag, c_{B,\downarrow,{\bf{k}}}^\dag)$ 
and $c_{\sigma,s,{\bf{k}}}^\dag$ are creation operators with $\vec{\sigma}$ and $\vec{s}$ Pauli matrices for the orbital (A,B) and spin degrees of freedom ($\uparrow, \downarrow$) in the $z$-direction respectively. Here, $k_0$, $\alpha$ and $\beta$ denote real model parameters. For $k_0>\beta$ , the model exhibits four Weyl nodes  at $Q_\gamma$ as shown in Fig.~\ref{bandstructure}(a) where we label $Q_{\gamma=1,2,3,4}$ as $(\beta,0,\sqrt{k_0^2-\beta^2})$, $(-\beta,0,-\sqrt{k_0^2-\beta^2})$, $(\beta,0,-\sqrt{k_0^2-\beta^2})$ and $(-\beta,0,\sqrt{k_0^2-\beta^2})$ respectively. We schematically 
depict the positions of the four Weyl nodes in Fig.~\ref{bandstructure}(b) choosing $k_{x}$-$k_{z}$ plane intersecting at $k_{y}=0$.

Using the transformation, $c_{{\bf{s}},{\bf{k}}}^{(\sigma)}=(c_{\sigma,\uparrow,{\bf{k}}}\pm c_{\sigma,\downarrow,{\bf{k}}})/\sqrt{2}$~\cite{Trauzettel_PRL} where ${\bf{s}}$=$(\uparrow, \downarrow)$ denotes the spin in the $x$-direction in this new basis, we can write the low energy Hamiltonian as the sum of four $2 \times 2 $ Hamiltonian near the four Weyl nodes as $\mathcal{H}_{W}=\sum_{\gamma=1}^{4}\sum^{\prime}_{\bf{k}}\Psi^{\dag}_{\gamma,{\bf{k}}}H_{\gamma}\Psi_{\gamma,{\bf{k}}}$, where $\Psi^{\dag}_{1,{\bf{k}}}=\Psi^{\dag}_{3,{\bf{k}}}=\left(c^{B \dag}_{\uparrow,{\bf{k}}},
c^{A\dag}_{\downarrow,{\bf{k}}}\right)$ and $\Psi^{\dag}_{2,{\bf{k}}}=\Psi^{\dag}_{4,{\bf{k}}}=(c^{A\dag}_{\uparrow,{\bf{k}}},c^{ B\dag}_{\downarrow,{\bf{k}}})$~\cite{Trauzettel_PRL}. Here, the 
$\sum^{\prime}_{\bf{k}}$ indicates the fact that crystal momenta $k_x, k_y, k_z$ are now restricted close to the Weyl points. 
The Hamiltonian $H_\gamma$ can be written as~\cite{Trauzettel_PRL}
\begin{eqnarray}
    H_{1,2} &=& (k_x \mp \beta)\textbf{\text{s}}_x+k_y \textbf{\text{s}}_y+(k_z \mp \sqrt{k_0^2-\beta^2})\textbf{\text{s}}_z\ , \nonumber \\
H_{3,4} &=& (k_x \mp \beta)\textbf{\text{s}}_x+k_y \textbf{\text{s}}_y-(k_z \pm \sqrt{k_0^2-\beta^2})\textbf{\text{s}}_z\ .
\label{lowenergyham}
\end{eqnarray}
Here, $k_y$ and $k_z$ are rescaled by $1/\alpha$ and $1/2\sqrt{k_{0}^2-\beta^2}$. The Weyl points $Q_{1}$ and $Q_{2}$ ($Q_{3}$ and $Q_{4}$) carry the positive (negative) chirality and form two 
time-reversed pairs.

We assume that the system remains metallic with an approximately linear dispersion even in the presence of disorder. 
This remains a good approximation as long as the Fermi energy is larger than the disorder induced gap in the system~\cite{Gregory2014}.
     Beyond that, we consider intra-orbital $s$-wave superconducting pairing 
which couples Weyl nodes of the same chirality. This is because at low energy, the inter-orbital pairing is 
known to be suppressed~\cite{Trauzettel2018}. Consequently, we can consider two decoupled superconducting Hamiltonians with opposite chirality as~\cite{Trauzettel2018}
\begin{eqnarray}
H^{+}_{S} &=\sum^{\prime}_{\bf{k}} \left(\Delta \,c_{1,\uparrow,{\bf{k}}}^\dag\, c_{2,\downarrow,-{\bf{k}}}^\dag + 1\leftrightarrow 2 \right) + \text{h.c.}, \label{hamSC1}\\
H^{-}_{S} &=\sum^{\prime}_{\bf{k}} \left(\Delta \,c_{3,\uparrow,{\bf{k}}}^\dag\, c_{4,\downarrow, -{\bf{k}}}^\dag + 3\leftrightarrow 4\right) + \text{h.c.}
\label{hamSC2}
\end{eqnarray}
where $c^\dag_{1,\uparrow,{\bf{k}}}$ and $c^\dag_{2,\downarrow,{\bf{k}}}$ in Eq.~(\ref{hamSC1}) [or  $c^\dag_{3,\uparrow,{\bf{k}}}$ and $c^\dag_{4,\downarrow, {\bf{k}}}$ in Eq.~(\ref{hamSC2})] are creation operators at Weyl points 1 and 2 (or 3 and 4) respectively. The pairing potential $\Delta$ couples electrons and holes only between the time-reversed pair of Weyl nodes \ie $Q_1$ to $Q_2$ and $Q_3$ to $Q_4$. Here, $\uparrow$ and $\downarrow$ represent the spin in $x$-direction.

\begin{figure}[h!]
    \begin{center}
        \includegraphics[width=3.5in,height=1.9in]{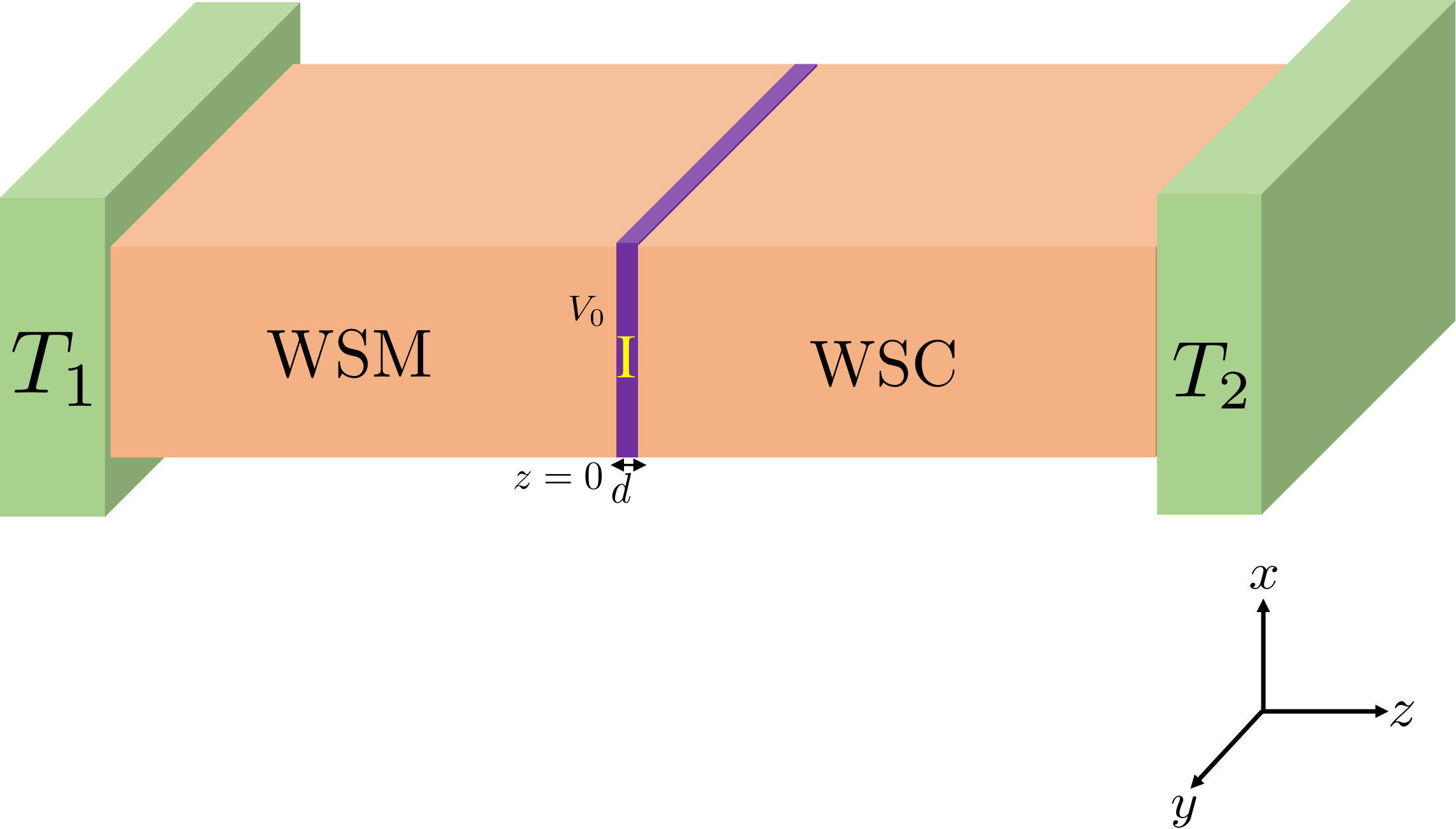}
    \end{center}
    \vspace{-0.2cm}
    \caption{Schematic diagram of our setup in which the WSM part is treated as the normal (N) region while the other side is a WSC with $s$-wave pairing (superconducting region S), thus forming a WSM-WSC heterostructure (orange). The green regions indicate the left and right reservoirs maintained at thermal equilibrium 
    with temperatures $T_{1}$ and $T_{2}$ respectively and at a finite voltage bias $eV$. A thin violet region symbolizes the insulating (I) barrier of width $d$ and height $V_{0}$,    
    sandwitched between the normal and superconducting regions. 
            }
            \label{NS}
\end{figure}

We analyse the heterostructure comprising  a WSM and a bulk WSC with $s$-wave pairing, considering the above mentioned model Hamiltonians. 
Our WSM-WSC setup is illustrated in Fig.~\ref{NS}. The thin violet region at the interface (see Fig.~\ref{NS}) indicates an insulating barrier with width $d$ and height $V_{0}$. We incorporate this to investigate thin barrier effects on thermal transport through the junction.   

Considering quasi one-dimensional transport along the $z$ direction, we can write the Bogoliubov-de Gennes (BdG) Hamiltonian for the positive chirality sector as an $8\times 8$ Hamiltonian 
\begin{eqnarray}
    \mathcal{H}_{+} = \left(\begin{array}{cc}
\large h^1_{\text{BdG}} & \emptyset \\[0.2cm]
\emptyset & \large h^2_{\text{BdG}}
\end{array}
\right)\ ,
\end{eqnarray}
where $\emptyset$ is a $4\times 4$ null matrix and each of the diagonal Hamiltonian can be written as
\begin{eqnarray}
h^1_{\text{BdG}} = \left(\begin{array}{cccc}
\quad \text{ Weyl node 1} & & \quad \,\, \Delta & 0 \\
 \quad(\text{electron space}) & & \quad \,\,0 & \Delta \nonumber \\[0.3cm]
 \hspace*{-0.8cm}\Delta^* & \hspace*{-0.7cm}0 & & \hspace*{-0.7cm}\text{ Weyl node 2} \nonumber \\
\hspace*{-0.8cm}0 & \hspace*{-0.7cm}\Delta^* & & \hspace*{-0.7cm}(\text{hole space})
\end{array}
\right) 
\end{eqnarray}
and
\begin{eqnarray}
h^2_{{\rm{BdG}}} = \left(\begin{array}{cccc}
\quad \text{ Weyl node 2} & & \quad \,\, \Delta & 0 \\
 \quad(\text{electron space}) & & \quad \,\,0 & \Delta \nonumber \\[0.3cm]
 \hspace*{-0.8cm}\Delta^* & \hspace*{-0.7cm}0 & & \hspace*{-0.7cm}\text{ Weyl node 1} \nonumber \\
\hspace*{-0.8cm}0 & \hspace*{-0.7cm}\Delta^* & & \hspace*{-0.7cm}(\text{hole space})
\end{array}
\right) \\
\end{eqnarray}
Here, $h^1_{\text{BdG}}$ and $h^2_{\text{BdG}}$ effectively represent the same BdG Hamiltonian. We focus only on $h^1_{{\rm{BdG}}}$ for the purpose of this paper. Using the Nambu basis in real space
$(\Psi_{1,\boldsymbol{\uparrow}}, \Psi_{1,\boldsymbol{\downarrow}}, \Psi_{2,\boldsymbol{\downarrow}}^\dag, -\Psi_{2,\boldsymbol{\uparrow}}^\dag)$, we can write $h^1_{\text{BdG}}$ as
\begin{eqnarray}
    h^1_{\text{BdG}} = \nu_z \left[-i \partial_r \cdot \textbf{\text{s}} - \mu(z) \, \textbf{\text{s}}_0\right] + \Delta(z) \, \nu_x \, \textbf{\text{s}}_0\ .
\end{eqnarray}
Here, Pauli matrices $\nu$ and \textbf{s} act on Weyl points and spin respectively. Considering the electronic transport along the $z$ direction, we define chemical potential $\mu$ and superconductong pairing potential $\Delta(z)$ for the geometry shown in Fig.~\ref{NS} as 
\begin{eqnarray}
\mu(z)&=&\mu_N \, \Theta(-z) +\mu_S\, \Theta (z)\ , \nonumber\\
\Delta(z)&=&\Delta(T)\, e^{i\alpha}\,\Theta(z)\ .
\end{eqnarray}

The temperature dependent superconducting gap is $\Delta(T)=\Delta_0 \tanh{(1.74 \sqrt{T_c/T-1})}$. $\Theta(z)$ is the Heaviside step function, $\alpha$ is the phase of the superconducting order parameter which is fixed at zero for simplicity and $T_c$ is the critical temperature of the superconductor. Here, $\mu_N$ and $\mu_S$ are the chemical potentials on 
WSM and WSC sides respectively. Furthermore, we introduce a unitary transformation in order to transfer the $k_{0}$ and $\beta$ dependence of the Hamiltonian to the 
wavefunction~\cite{Trauzettel2018}.

Similarly, one can write the BdG Hamiltonian for negative chirality sector ($H_{-}$) by replacing $\partial_z$ by $-\partial_z$. We focus only on the positive chirality sector in this work and it is straightforward to extend the thermal transport analysis for the negative chirality Weyl fermions.

The mean-field theory of superconductivity demands that $\mu_S\gg\Delta_0$, or equivalently, that the superconducting coherence length $\xi$ is much larger than the Fermi-wavelength in the 
superconducting region~\cite{SARBeenakker}. Note that, we fix $\mu_S=100\Delta_0$ to satisfy this requirement throughout our analysis. 

\begin{figure*}[]
    \begin{center}
        \includegraphics[width=7.2in,height=2.8in]{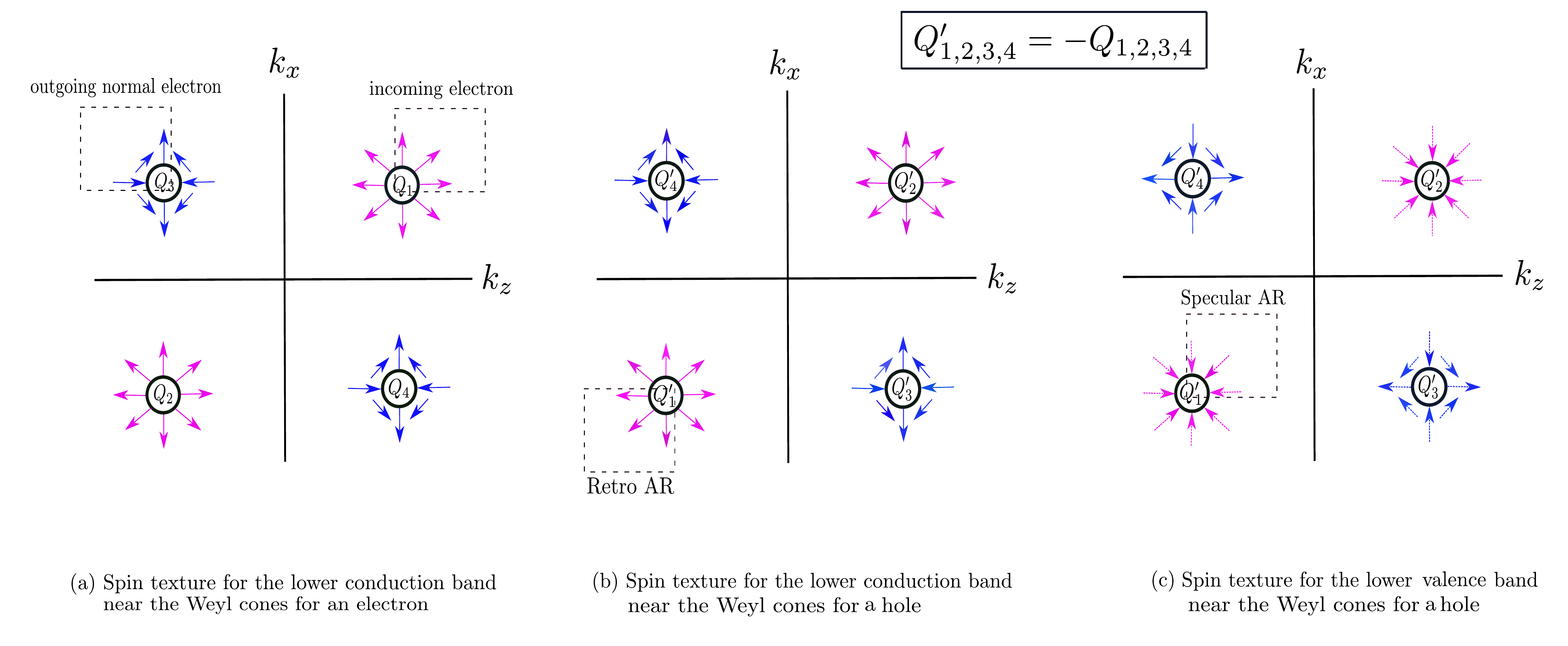}
    \end{center}
              \vspace{-0.5cm}
    \caption{\hspace*{-0.1cm} (a) The spin texture is shown for the lower conduction band near the Weyl cones for an electron. The same is shown 
    for the lower conduction band and lower valence band near the Weyl cones for a hole in panels (b) and (c) respectively. We choose the other model parameters as 
    $\beta=1$, $k_0=\sqrt{2}$ and $k_{y}=0$.}
    \label{spintexture}
\end{figure*}

To understand the nature of normal and Andreev reflection (AR) at the interface between WSM and WSC, it is necessary to analyse the spin texture around the Weyl nodes for both electrons and holes. 
For this, we employ the low energy Hamiltonians [Eq.~(\ref{lowenergyham})] to find the expectation values of spin vector for the lower conduction band around each Weyl point and show them in 
Fig.~\ref{spintexture}. It is evident from Fig.~\ref{spintexture}(a) that the conduction electron incident from Weyl point $Q_1$ is reflected as a normal electron from the opposite Fermi surface at 
$Q_3$ to preserve the spin quantum number. 
The hole due to retro-AR comes from the time reversed partner which is at $Q_1' = -Q_1$ and carries opposite texture compared to the incoming electron.This is schematically shown in 
Fig.~\ref{spintexture}(b). Specular AR takes place from the valence band of the hole, and this is depicted in Fig.~\ref{spintexture}(c).
\vspace{+0.5cm}	
\subsection{Scattering matrix approach}
Due to the superconducting pairing between the time-reversed pairs of Weyl nodes with the same chirality~\cite{Trauzettel2018}, we can separately compute contributions arising
from the positive and negative chirality Weyl nodes. Considering the positive chirality Weyl nodes in our study, we investigate the thermal transport employing the scattering matrix approach~ 
\cite{BTK,Bagwell1993,Arijit2017} for the non-interacting electrons in WSM-WSC setup
(see Fig.~\ref{NS}). Let us consider the electronic transport along the perpendicular direction to the interface between normal and superconducting part of the WSM, which means that the momentum parallel to the interface $k_{||}\,(=k_x,k_y)$ is conserved. We consider the excitation energy $E>0$, such that the incoming electron is above the Fermi level at energy $\mu_N+E$ while the Andreev reflected hole is at energy $\mu_N-E$.

Using BdG Hamiltonian for the WSM-WSC junction, the wave-functions in the normal side (WSM) for a given excitation energy $E$ can be written as
\begin{eqnarray}
\psi_{\overrightarrow{e}} (z) &=& \left(\cos\varphi_e,e^{i\theta_k} \sin\varphi_e,0,0\right)^T e^{i k_ez}\ , \nonumber\\
\psi_{\overleftarrow{e}} (z) &=& \left(e^{-i\theta_k} \sin\varphi_e,\cos\varphi_e,0,0\right)^T e^{-i k_e z}\ , \nonumber\\
\psi_{\overrightarrow{h}} (z) &=& \left(0,0,-e^{-i\theta_{k}} \sin\varphi_h,\cos\varphi_h\right)^T e^{i k_{h} z}\ , \nonumber\\
\psi_{\overleftarrow{h}} (z) &=& \left(0,0,\cos\varphi_h,-e^{i\theta_k} \sin\varphi_h\right)^T e^{-i k_h z}\ ,
\label{wf_normal}
\end{eqnarray}
\noindent
where $k_{(e,h)}\!=\!\sqrt{(E \pm \mu_N)^2-k_{||}^2}$ are the momenta of the electron and hole respectively, $\varphi_{(e,h)}=\tan^{-1}(k_{||}/k_{(e,h)})/2$ are the incident angles of the incoming electron 
and hole respectively and $\theta_{k}=\tan^{-1}(k_y/k_x)$.

The wave functions in the WSC region can be written as

\begin{widetext}
        \begin{eqnarray}
\Psi_{\overrightarrow{eq}} (z) &=& \left(e^{i\beta}\cos\phi_e,e^{i\beta}\,e^{i\theta_{k}} \sin\phi_e,e^{-i\alpha}\cos\phi_e,e^{-i\alpha}e^{i\theta_{k}}\sin\phi_e\right)^T e^{iq_e z}\ , \nonumber\\
    \Psi_{\overleftarrow{eq}} (z) &=& \left(e^{i\beta}e^{-i\theta_{k}} \sin\phi_e,e^{i\beta}\cos\phi_e,e^{-i\alpha}e^{-i\theta_{k}} \sin\phi_e,e^{-i\alpha}\cos\phi_e\right)^T e^{-iq_e z}\ , \nonumber\\
    \Psi_{\overrightarrow{hq}} (z) &=& \left(e^{i\alpha}e^{-i\theta_{k}}\cos\phi_h,e^{i\alpha}\sin\phi_h,e^{i\beta}e^{-i\theta_{k}}\cos\phi_h,e^{i\beta}\sin\phi_h\right)^T e^{iq_h z}\ , \nonumber\\
    \Psi_{\overleftarrow{hq}} (z) &=& \left(e^{i\alpha}\sin\phi_h,e^{i\alpha}\,e^{i\theta_{k}} \cos\phi_h,e^{i\beta}\sin\phi_h,e^{i\beta}e^{i\theta_{k}}\cos\phi_h\right)^T e^{-iq_hz}\ ,\nonumber\\
    \end{eqnarray}
    \vspace{-0.5cm}
\end{widetext}
\noindent
where, $q_{(e,h)}=\sqrt{(\mu_S\pm \Omega)^2-k_{||}^2}$ and $\phi_{(e,h)}=\tan^{-1}(k_{||}/q_{(e,h)})/2$ are the momenta (close to Weyl points) and incident angles 
of the electron and hole like quasiparticles respectively inside the superconducting region. 
For $E\leq \Delta$ (sub-gapped regime), $\beta=\cos^{-1}(E/\Delta)$ and $\Omega=i\sqrt{\Delta^2-E^2}$ while for $E>\Delta$ (above the gap), $\beta=-i\,\cosh^{-1}(E/\Delta)$ and 
$\Omega=\sqrt{E^2-
\Delta^2}$. Considering the scattering between the same chirality Weyl nodes, we can write the scattering states on both sides of the junction as
\begin{eqnarray}
\Psi_{<} (z) &=& \psi_{\overrightarrow{e}} (z) +r_e\, \psi_{\overleftarrow{e}} (z) + r_h\, \psi_{\overleftarrow{h}} (z)\ , \nonumber \\
\Psi_{>} (z) &=& t_e \, \Psi_{\overrightarrow{eq}} (z) + t_h \, \Psi_{\overleftarrow{hq}}\ ,
\label{scatcoef}
\end{eqnarray}
where, $r_{e,h}$ corresponds to the normal and AR scattering coefficients respectively, while $t_{e,h}$ denotes the transmission coefficients for the electron and hole like quasiparticles respectively. 
We obtain the $4\times 4$ scattering matrix by matching the wave functions [Eq.~(\ref{scatcoef})] on both sides of the junction at $z=0$ ($\Psi_{<}(z)=\Psi_{>}(z)\rvert_{z=0}$). 

\section{Thermal transport}
\label{thermaltransport}
For the non-interacting electrons, the total transmission function for the transport of the electronic charge through the junction can be computed using the scattering matrix approach. 
Integrating over the transverse modes one can write~\cite{Arijit2016,Linder2008}
\small
\begin{eqnarray}
\mathcal{T} (E)=\int \frac{d^2 k_{||}}{4 \pi^2} \left( 1-R_e(E)+ \text{Re}\!\left[ \frac{\cos 2\varphi_h}{\cos 2\varphi_e} \right]\, R_h(E)\right)\ ,
\end{eqnarray}
\normalsize
where, $k_{||}$ is the momentum parallel to the junction interface, $R_e$ and $R_h$ are the normal and AR probability respectively, $\varphi_e$ and $\varphi_h$ are the angle 
of incidence for electron and hole respectively. Changing the integration variable $k_{||}$ to $\varphi_e$ using $k_{||}=(E+\mu_N)\sin 2\varphi_e$, we can write
\begin{align}
\mathcal{T} (E) &= \frac{1}{2\pi} \int d\varphi_e \left( 1-R_e(E) + \text{Re} \left[\frac{\cos 2\varphi_h}{\cos 2\varphi_e} \right] \, R_h(E)\right) \nonumber\\
&\hspace*{4cm}(E+\mu_N)^2 \sin 4\varphi_e\ .
\end{align}
\normalsize
While the electrical charge transport can distinguish between the electron and hole contribution by the sign of $R_e$ and $R_h$, the thermal transport treats both the electron 
and the hole as a particle carrying the number density along the same direction (see Appendix~\ref{NS_tempbias} for details). Consequently, for the thermal transport, the 
transmission function for the energy transport through the junction can be computed using~\cite{Linder2008,Arijit2016} as
\begin{align}
\mathcal{T}_{\text{therm}} (E) &= \frac{1}{2\pi} \int d\varphi_e \left( 1-R_e(E) - \text{Re} \left[\frac{\cos 2\varphi_h}{\cos 2\varphi_e} \right] \, R_h(E)\right) \nonumber\\
&\hspace*{4cm}(E+\mu_N)^2 \sin 4\varphi_e\ .
\end{align}
\subsection{Electrical conductance and Thermoelectric coefficient}
We begin with the general expression for total electrical current through the normal-superconductor (NS) junction in the presence of voltage and temperature biases~\cite{BTK} as
\begin{align}
I &= \frac{2e}{h} \int_{-\infty}^\infty \Big[f_N(E-e\Delta V,T+\Delta T)-f_S(E,T)\Big] \nonumber\\
&\hspace*{5cm} \mathcal{T} (E)\, dE\ .
\label{eq_BTK}
\end{align}
\noindent
where, $f_{N}$ and $f_{S}$ are the Fermi distribution functions of the two reservoirs to which the NS setup is attached.
 
In the linear response regime, one can expand the Fermi functions for very small voltage and temperature differences as 
\begin{eqnarray}
f_N(E-e\Delta V,T+\Delta T)&=&f_0(E,T)-e\Delta V \frac{\partial f_0}{\partial E}+\Delta T \frac{\partial f_0}{\partial T}\ , \nonumber \\
f_S(E,T)&=&f_0(E,T)\ .
\label{Ffunction}
\end{eqnarray}
where $f_0$ is the average Fermi distribution at energy $E$ and temperature $T$. Using Eq.~(\ref{Ffunction}) we can write the total current as (see Appendix~\ref{NS_tempbias} for the complete  
derivation following Ref.~\cite{BTK})
\begin{eqnarray}
I=\frac{2e}{h} \int_{-\infty}^{\infty} \left[-e\Delta V \frac{\partial f_0}{\partial E} +\Delta T \frac{\partial f_0}{\partial T} \right] \mathcal{T} (E)\,dE\ .
\label{current}
\end{eqnarray}

Hence, comparing the total current with the Onsager relation 
\begin{eqnarray}
I=G\Delta V+L_{12}\Delta T\ ,
\label{Onsager}
\end{eqnarray}
we arrive at the expressions for electrical conductance and the thermoelectric coefficient~\cite{Bagwell1993,Sivan_Imry1993} as
\begin{eqnarray}
G=\frac{2 e^2}{h} \int_{-\infty}^{\infty} \left(-\frac{\partial f_0}{\partial E}\right) \mathcal{T} (E)\, dE\ , \nonumber \\
L_{12}=-\frac{2 e}{h} \int_{-\infty}^{\infty} \left(-\frac{\partial f_0}{\partial T}\right)   \mathcal{T} (E) \,dE\ .\nonumber \\
\end{eqnarray}

\subsection{Thermal conductance, Thermopower and Figure of merit}
The normalized thermal conductance through the NS interface can be defined as~\cite{Bagwell1993} 
\begin{eqnarray}
\kappa/\kappa_0=\int_{-\infty}^\infty (E-\mu_N) \left(-\frac{\partial f_0}{\partial T} \right)  \mathcal{T}_{\text{therm}} (E)\, dE\ .
\label{thermcond}
\end{eqnarray}
where, $\kappa_0$ is the thermal conductance due to ballistic channels in the bulk at energy $(E+\mu_N)$ and $T=T_c$.  
This is given by 
\begin{eqnarray}
\kappa_0= \int_{-\infty}^\infty \, dE\, \frac{(E+\mu_N)^2}{4\pi} (E-\mu_N) \left(\frac{\partial f_0}{\partial T} \right)_{T=T_c}\ .
\end{eqnarray}

When a temperature gradient is maintained in a metal and no electric current is allowed to flow, 
a steady-state electrostatic potential difference builds up between the high and low temperature regions. This is called the thermoelectric effect and is measured by the thermopower 
(the Seebeck coefficient) which by definition is given by~\cite{Ashcroft76}
\begin{eqnarray}
S=-\left(\frac{\Delta V}{\Delta T}\right)_{I=0}\ .
\end{eqnarray}
Using Eq.~(\ref{Onsager}) the analytical expression for the thermopower can be written as 
\begin{eqnarray}
S=\frac{L_{12}}{G}\ .
\label{eq_thermpower}
\end{eqnarray}
where, $S$ is measured in units of $k_{B}/e$.

The performance of a thermoelectric device is characterized by a dimensionless quantity called the figure of merit,
\begin{eqnarray}
zT=\frac{S^2 G T}{\kappa} \ .
\label{eq_Fom}
\end{eqnarray}
The major objective is to attain higher values of figure of merit in different materials/hybrid junctions, that would be desirable for their potential applicability in thermoelectric devices. It is clear from 
Eq.~(\ref{eq_Fom}) that this can be achieved by improving electrical transport properties and reducing thermal conductivity.

\subsection{Wiedemann-Franz law}
The Wiedemann-Franz (WF) law states that the ratio of thermal conductance, $\kappa$ to electrical conductance, $G$ is proportional to the absolute temperature 
\begin{eqnarray}
\frac{\kappa}{G}=\mathcal{L}\, T\ ,
\label{WFlaw}
\end{eqnarray}
where the proportionality constant $\mathcal{L}$ is a universal number called the Lorentz number. For an ideal Fermi gas, we have $\mathcal{L}=\mathcal{L}_0=\frac{\pi^2}{3}\left(\frac{k_B}{e} 
\right)^2$. 
Violation of the WF law has been reported for semiconductors~\cite{Goldsmid1956}, zero-gap systems \eg graphene~\cite{Crassno2016,Marzari2016}, quasi one-dimensional Luttinger 
liquids~\cite{Wakeham2011} and heavy-fermion materials~\cite{Tanatar2007}. We explore the validity of the WF law in our WSM-WSC setup and unveil several interesting features near the Weyl point 
which we discuss in the next section. 

\section{Numerical results}\label{numerical results}

In this section, we proceed to discuss our numerical results for thermal transport through the WSM-WSC setup. We compute the scattering probabilities employing Eq.~(\ref{scatcoef}) and find the
relevant physical quantities using the analytical expressions discussed in Sec.~\ref{thermaltransport}.

\subsection{Thermal conductance}
\begin{figure}[H]
    \begin{center}
        \includegraphics[width=3.4in,height=1.5in]{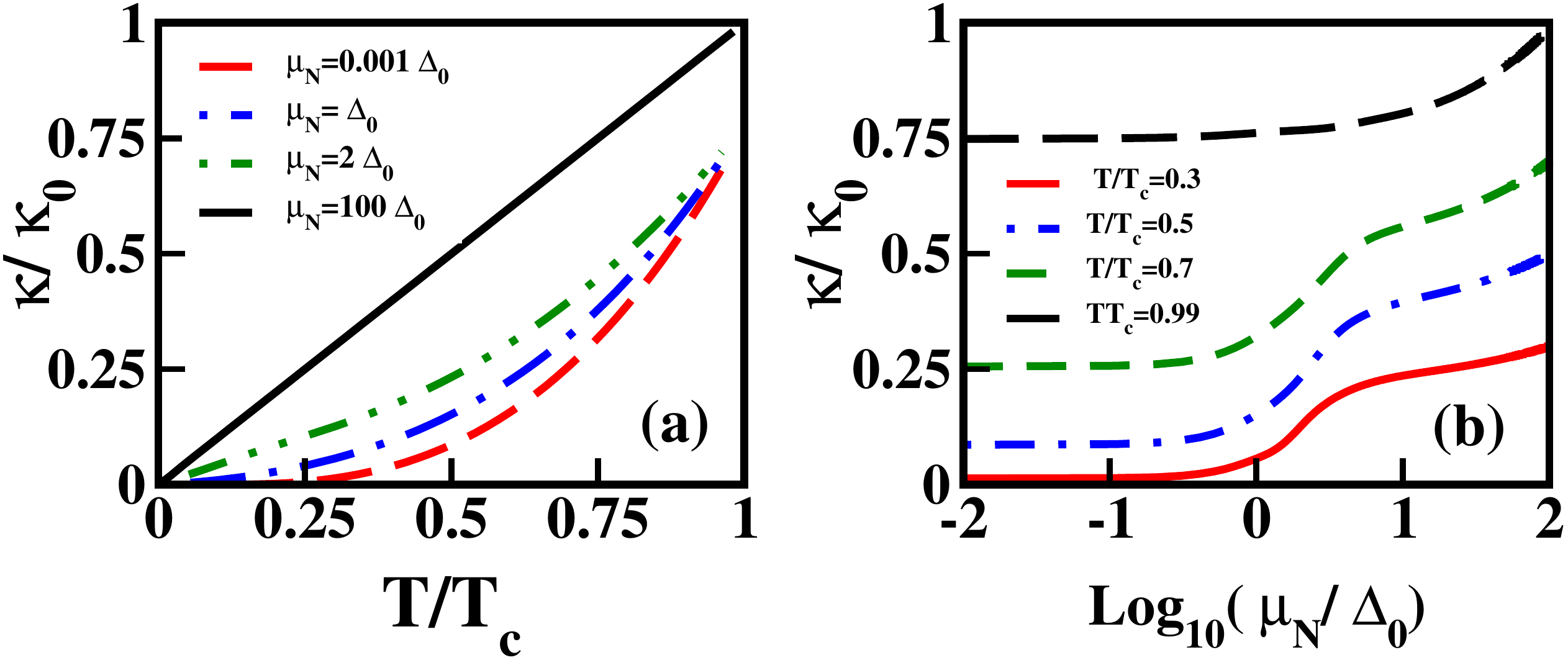}
    \end{center}
    \caption{\hspace*{-0.1cm} (a) The behavior of normalized thermal conductance is shown as a function of temperature for $\mu_N/\Delta_0=0.001,\, 1,\, 2,\, 100$ represented by 
    dashed, 
    dot-dashed, double dot-dashed and solid line respectively. (b) The normalized thermal conductance with respect to the logarithm of $\mu_{N}/\Delta_{0}$ is depicted for $T/T_{c}=0.3, 0.5, 0.7~ 
    \rm {and}~0.99$ represented by longer dashed, dashed, dot-dashed and solid lines respectively.}
    \label{thermcondplot}
\end{figure}

We compute the normalized thermal conductance using Eq.~(\ref{thermcond}) and show its behavior as a function of temperature in Fig.~\ref{thermcondplot}(a). 
For $\mu_N<\Delta_0$, the thermal conductance remains vanishingly small for lower temperatures and starts increasing exponentially with the enhancement of temperature. 
The zero value of thermal conductance at $T=0$ is due to the perfect AR in the subgapped regime, which blocks the thermal transport. When the temperature is very small ($T\ll T_{c}$), 
electrons with energy at least $\mu_N$ contribute to the thermal conductance. We show this in Fig.~\ref{thermcondplot}(a) for $\mu_N/\Delta_0=1$ and $\mu_N/\Delta_0=2$.
    Clearly, the thermal conductance increases as we enhance $\mu_N$, due to the availability of more states. The blockage of thermal transport (at $T=0$) continues as long as the transport is dominated 
    by the superconducting Cooper pairs. Further, thermal conductance exhibits exponential rise for $\mu_N\leq\Delta_0$ as one approaches towards $T_{c}$.
    We attribute this behavior to the spin singlet $s$-wave superconductor in Weyl NS junction as has been reported earlier in superconducting hybrid junctions based on Dirac like materials~
    \cite{Linder2008,Arijit2016, Arijit2017}. When the chemical potential on the normal WSM side becomes much larger than the superconducting gap, the NS junction becomes transparent and behaves 
    like a metal. The metallic behaviour of the thermal conductance is known to be linear with the increase of temperature~\cite{Averback2005,Rashedi2011}, and we find this characteristic for $\mu_N/
    \Delta_0=100$ as shown in Fig.~\ref{thermcondplot}(a). This is because at $\mu_N\gg\Delta_0$, the incoming electrons easily overcome the superconducting gap on the right side which results in a 
    linear increase in thermal conductance.

  Further, we discuss the behavior of thermal conductance as a function of chemical potential $\mu_N$ in the normal side, at a fixed temperature. To show the clear dependence on $\mu_N$ at a larger 
  scale (0.01$\Delta_0$ to 100$\Delta_0$) we demonstrate this using $\log$ scale for $\mu_N$ in Fig.~\ref{thermcondplot}(b). It is evident that, as we increase the temperature towards $T_{c}$, there is a 
  larger enhancement in the thermal conductance as we move away from $\Delta_0$.

\subsection{Wiedemann-Franz law}
\label{WF}

\begin{figure}[H]
    \begin{center}
        \includegraphics[width=3.4in,height=2.4in]{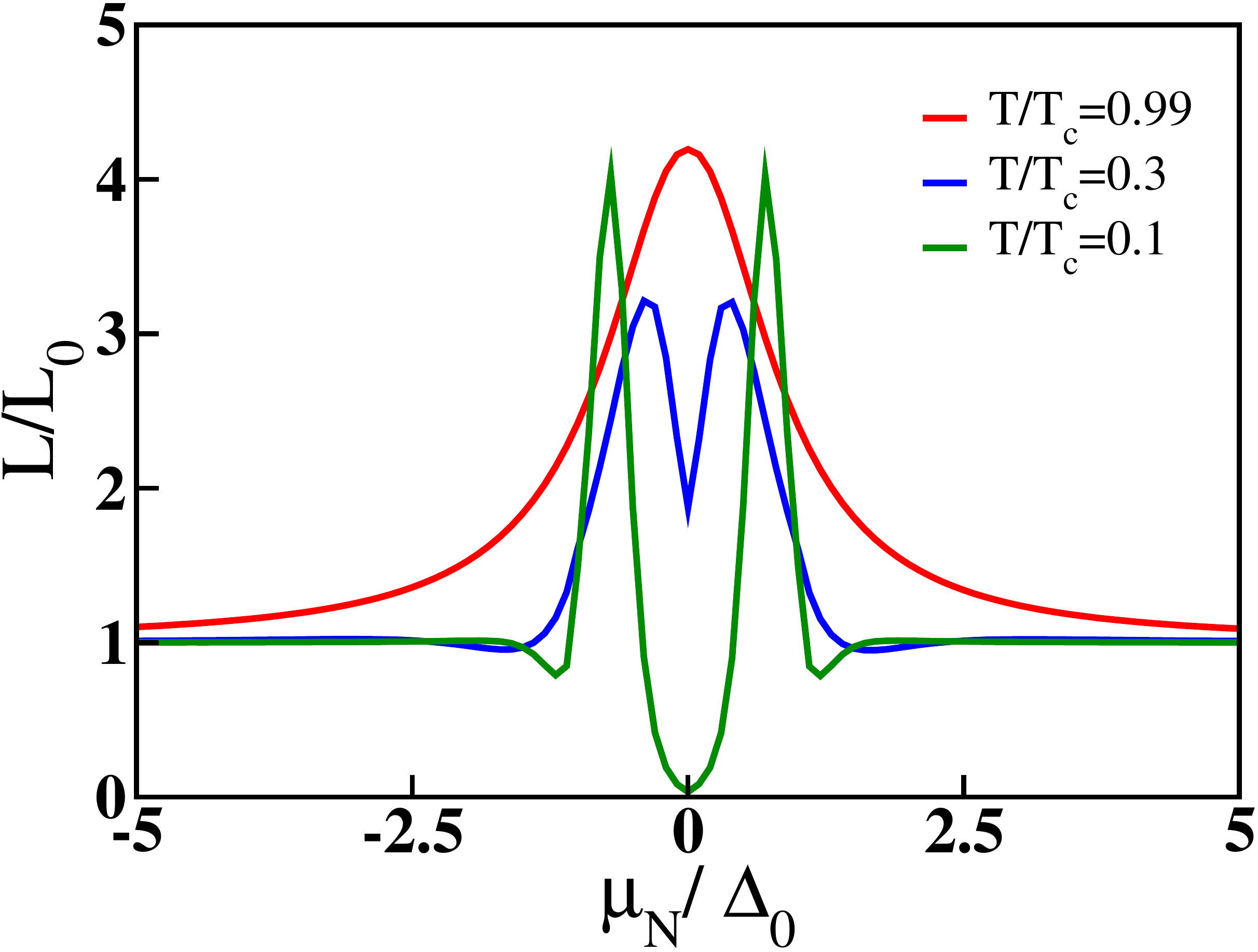}
    \end{center}
    \caption{\hspace*{-0.1cm} The variation of Lorentz number, in the WSM-WSC junction, is shown as a function of $\mu_N$ for three different values of temperatures. 
    Red, blue and green colour denote the corresponding behavior for $T/T_c=0.99, 0.3, 0.1$ respectively.}
    \label{WFlawplot}
\end{figure}
    We discuss the features of the Weidemann-Franz law in this subsection. We numerically compute the thermal and electrical conductance to find the Lorentz number (in terms of Lorentz number $L_0$ 
for a metal), as given in Eq.~(\ref{WFlaw}). We depict the Lorentz number with the variation of the chemical potential in the WSM side of the junction for three values of temperature 
$T/T_c=0.99,\, 0.3,\, 
0.1$ in Fig.~\ref{WFlawplot}. The symmetric behaviour of the Lorentz number with respect to the chemical potential $\mu_N$ is due to the particle-hole symmetry of the system. 
    We generally arrive at the larger Lorentz number when there is a small contribution to the thermal conductance. For moderate temperature (\eg $T/T_c=0.3$ in Fig.~\ref{WFlawplot}), the Lorentz 
    number is 
    finite, even at $\mu_N=0$. This is due to the reduction of the superconducting gap with the increase of temperature, and we expect that this is responsible for finite thermal conductance even at 
    $\mu_N<\Delta_0$. Note that, there is always a dip in the Lorentz number at the Weyl point ($\mu_N=0$) below $T_c$. When $T/T_{c}=0.1$, this dip becomes maximum and touches the Weyl point 
    due to the unavailibility of density of states. However, at $T/T_{c}=0.99$ the superconducting gap becomes vanishingly small and we find the Lorentz number for the system to be 4.1958 at the 
    Weyl point. 
    This large value can be atrributed to the quasi-particle states close to the superconducting gap. We analytically compute the Lorentz number for a bulk WSM (see Appendix~\ref{WF_law2} for details) 
    following the semi-classical approach and we find it to be 5.13. This is of the same order of magnitude as the numerical estimate here, where we consider the quantum behaviour of the system. 
\begin{figure}[H]
    \begin{center}
        \includegraphics[width=3.5in,height=1.6in]{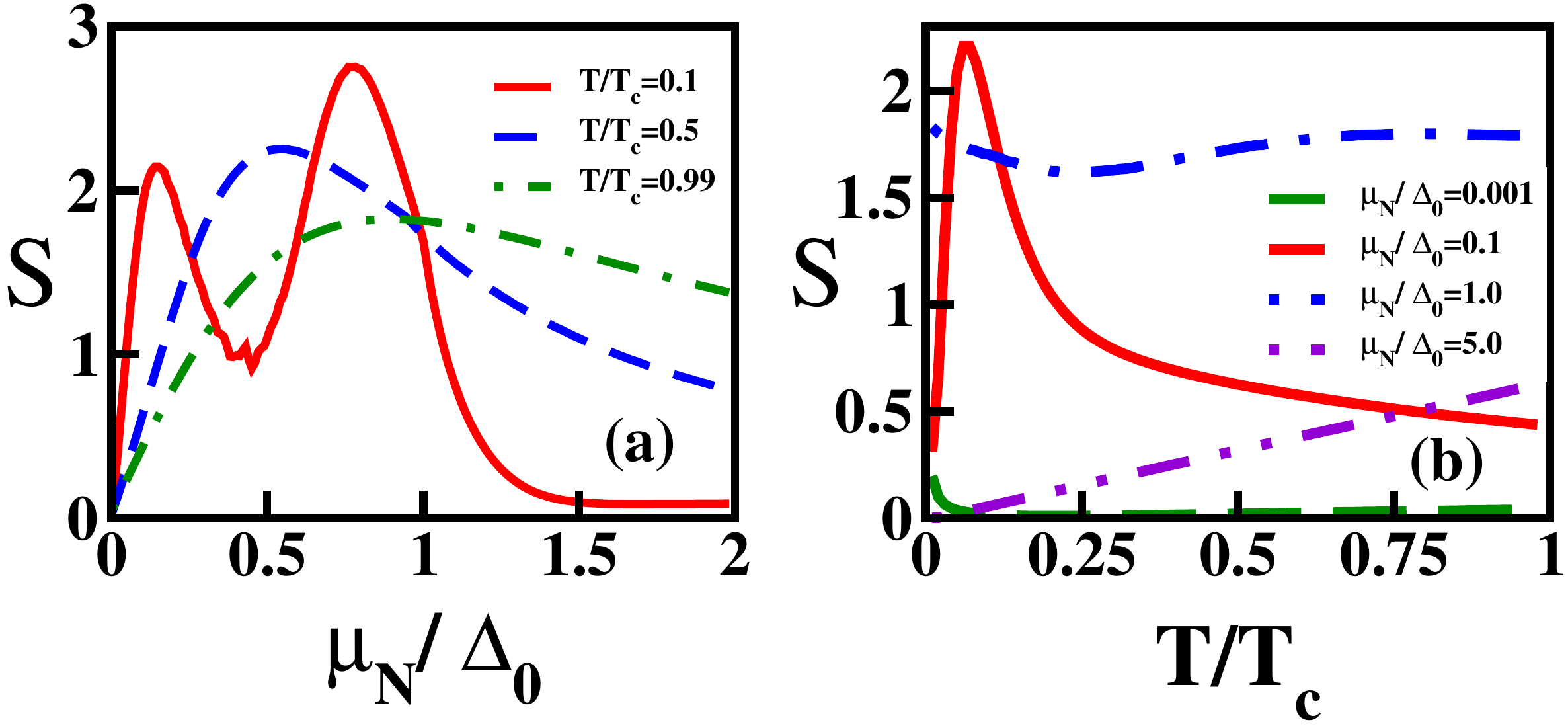}
    \end{center}
    \caption{\hspace*{-0.1cm} The behavior of thermopower or Seeback coefficient (in unit of $k_B/e)$ is depicted as a function of (a) $\mu_N/\Delta_{0}$ for $T/T_c=0.1,\, 0.5,\, 0.99$ 
    represented by red (solid), blue (dash) and green (dot-dash) lines respectively and (b) $T/T_{c}$ for $\mu_N/\Delta_0=0.001,\,1,\,2,\,100$ represented by green (dash), red (solid), blue (dot-dash) 
    and purple (double dot-dash) lines respectively.}
    \label{fig_thermpower}
\end{figure}

The physical reason behind this peculiar behaviour (violation of the WF law) of the Lorentz number within the range $-\Delta_0<\mu_N<\Delta_0$, can be attributed to the temperature 
dependence of the 
superconducting gap in our NS Weyl junction. Such deviations of the WF law from $L_0$ (close to the Dirac point) have been reported earlier in case of graphene~\cite{Crassno2016,Marzari2016}. 
    Beyond this region, the Lorentz number converges to $L_0$ (Lorentz number for metals) for all temperatures. For $\mu_N>\Delta_0$, the transport occurs via the states above the sub-gap energy.  
    This means that the WSM-WSC junction becomes transparent to electronic transport, and once again, we witness metal-like behaviour as $\mu_{N}\gg \Delta_{0}$.

\subsection{Thermopower and figure of merit}
\label{sec_thermpowr}

\begin{figure}[H]
    \begin{center}
        \hspace*{0cm}\includegraphics[width=3.4in,height=1.5in]{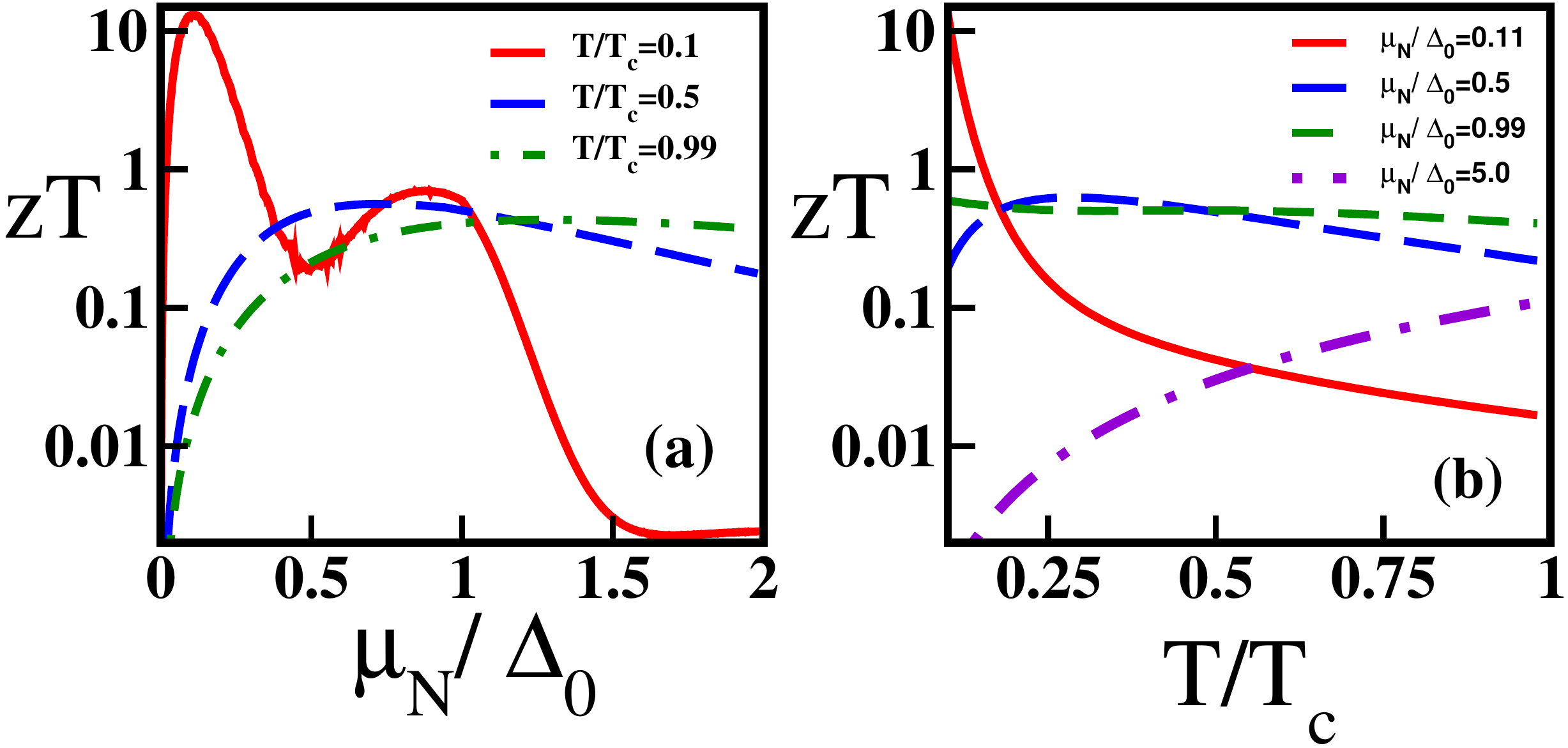}
    \end{center}
    \caption{\hspace*{-0.1cm}(Color online) The figure of merit $zT$ is depicted in the $\rm \log$ scale as a function of (a) $\mu_N/\Delta_{0}$ for $T/T_c=0.1,\,0.5,\,0.99$ represented by red (solid), 
    blue (dash) and green (dot-dash) lines respectively and (b) $T/T_{c}$ for $\mu_N/\Delta_0=0.11,\,0.5,\,0.99,\,5$ denoted by red (solid), blue (large dash), green (small dash) and purple (dot-dash) 
    lines respectively.}
    \label{fig_fom}
\end{figure}

Here, we analyse the thermopower or Seebeck coefficient that also describes the thermoelectric properties of the system. 
We have defined this quantity in Sec.~\ref{thermaltransport} (see Eq.~(\ref{eq_thermpower})). We present the observed characteristics of thermopower 
in Fig.~\ref{fig_thermpower} as a function of (a) the doping level $\mu_N$ in the WSM side and (b) the temperature $T$. Note that the thermopower at $T\ll T_c$ can be clearly distinguished
from the behaviour at $T<T_c$ for all values of $\mu_N$.

The efficiency of converting heat into electricity is related to the thermoelectric figure of merit, $zT$ defined in Sec.~\ref{thermaltransport}. We compute the dimensionless figure of merit 
employing $S^2GT/\kappa$ (Eq.~(\ref{eq_Fom})) and show the corresponding behavior in Fig.~\ref{fig_fom}(a) as a function of $\mu_N/\Delta_0$ at fixed temperatures. Interestingly, we find larger 
values of $zT$ for lower temperature ($T\ll T_{c}$) and doping ($\mu_{N}\ll \Delta_{0}$), as shown in Fig.~\ref{fig_fom}(a). 
Quantitatively, such high values of $zT$ arises due to small values of $\kappa$ and large $S$ in this regime. However, we do not fully understand the physical reason behind having such a large value 
of $zT$. Nevertheless, we note that the two-dimensional Dirac materials, whose low energy dispersion is effectively described by the massless Dirac Hamiltonian, have been shown to maximize 
$zT$ upto 15~\cite{Nugraha2019}. The behaviour of $zT$ is also shown in Fig.~\ref{fig_fom}(b) as a function of temperature for fixed $\mu_N$. Note that, high values of $zT\sim 9-10$ also appear 
for low temperatures when $\mu_N/\Delta_0\ll 1$.

\subsection{Effect of Insulating Barrier}
\label{insulatingb}
    \begin{figure}[H]
    \begin{center}
        \includegraphics[width=3.4in,height=1.5in]{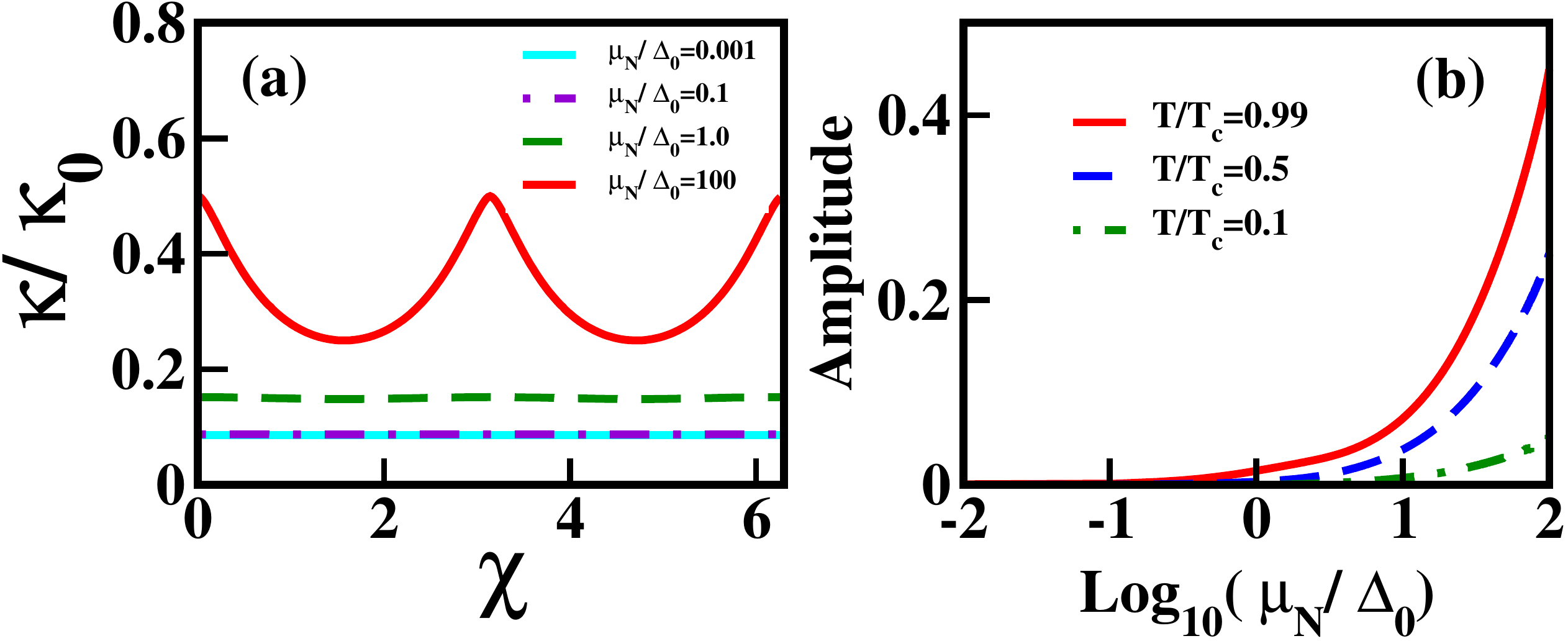}
    \end{center}
    \caption{\hspace*{-0.1cm}(Color online) (a) The normalized thermal conductance is depicted as a function of the barrier strength $\chi$ at $\mu_N/\Delta_0=0.001,\, 0.1,\, 1,\, 100$, $T/T_{c}=0.5$ 
    (b) the maximum amplitude of the thermal conductance oscillations as a function of $\mu_{N}$, is shown in the $\rm \log$ scale 
    at $T/T_c=0.99,\, 0.5,\, 0.1$.}
    \label{fig_barrier_thermcond}
\end{figure}

In this subsection, we investigate the changes in thermoelectric properties 
caused by an insulating barrier introduced at the WSM-WSC interface (see Fig.~\ref{NS}).
We assume the barrier to be of length $d$ and height $V_0$. This can be modelled as $U(z)=V_{0}~\Theta(z)\Theta(d-z)$. 
We carry out our analysis by matching the wave function at the two interfaces, assuming one to be at $z=0$ and the other at $z=d$. The wave function in the insulating region can be written as 
Eq.~(\ref{wf_normal}) with $\mu_N$ replaced by $\mu_N-V_0$. Here, we consider the thin barrier limit.
In this limit, one can define a finite quantity $\chi=V_0 d$~\cite{Arijit2016} with $V_0\rightarrow \infty$ and $d\rightarrow0$, called the 
strength of the barrier. We present here the distinctive behaviour of the thermoelectric properties with respect to this barrier strength $\chi$.

To begin with, we show the behavior of thermal conductance with the variation of $\chi$ in Fig.~\ref{fig_barrier_thermcond}(a). 
Clearly, $\pi$-periodic oscillations in $\kappa$ are found for larger doping ($\mu_N/\Delta_0=100$). On the other hand, the amplitude of oscillations becomes negligibly small  
as one decreases $\mu_{N}$. We also find that the maximum amplitude (difference between $\lvert \kappa(\chi=0)-\kappa(\chi=\pi/2)\rvert$) of oscillations increases 
as we increase the value of $\mu_N$ at any fixed temperature. We present this behavior in Fig.~\ref{fig_barrier_thermcond}(b). 
The maximum amplitude of oscillations increases more rapidly as one approaches $T\rightarrow T_{c}$ for $\mu_{N}\gg \Delta_{0}$. 

\begin{figure}[H]
    \begin{center}
        \includegraphics[width=3.4in,height=1.5in]{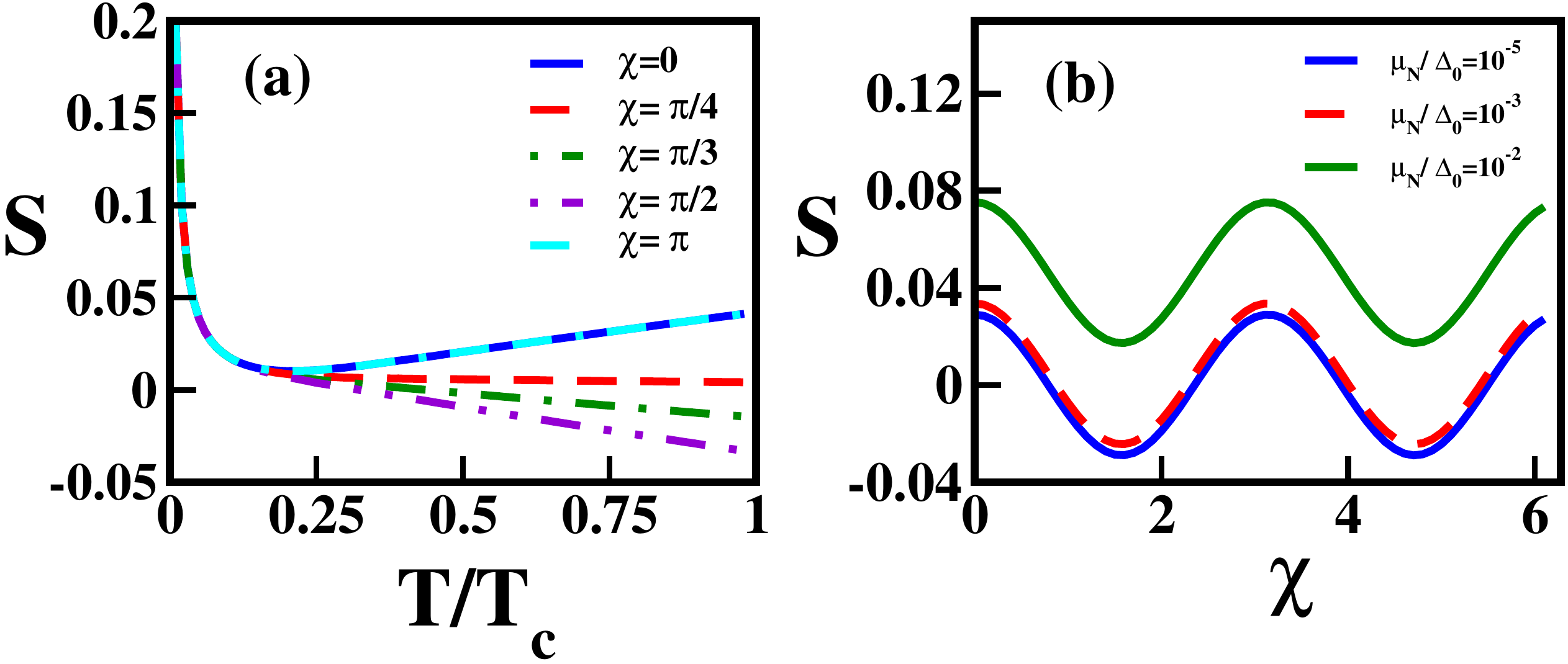}
    \end{center}
    \caption{\hspace*{-0.1cm}(Color online) (a) Thermopower $S$ is shown as a function of temperature at $\chi=0,\, \pi/4,\, \pi/3,\, \pi/2,\, \pi$. 
Here we choose $\mu_N/\Delta_0=0.001$. (b) $S$ is depicted with the variation of 
insulating barrier strength $\chi$ at $\mu_N/\Delta_0=10^{-5},\,10^{-3},\,10^{-2}$ and $T/T_c=0.8$.}
    \label{fig_barrier_thermpower}
\end{figure}

Afterwards, we study thermopower $S$ with the variation of $T/T_{c}$ for diferent values of $\chi$ 
and show its corresponding behavior in Fig.~\ref{fig_barrier_thermpower}(a). 
Note that the thermopower changes sign for $\chi=\pi/3, \pi/2$. This characteristic feature of changing sign in thermopower is also clearly visible in Fig.~\ref{fig_barrier_thermpower}(b) where
we illustrate its behavior with respect to $\chi$ for different values of $\mu_{N}\ll \Delta_{0}$. 
We expect that this sign change in $S$ takes place due to the electrons close to the Weyl point. 
\vspace{0.2cm}

\section{Summary and discussion}
\label{Summary and Discussion}
To summarize, in this work, we have investigated thermal transport through normal-superconductor junctions of inversion symmetry broken Weyl semimetals (WSM-WSC heterostructures). 
We have discussed the characteristic features of thermal conductance, thermopower and figure of merit under different sets of parameter values. We have shown that the normalized thermal 
conductance has the conventional exponential dependence on the temperature for lower dopings (chemical potential $\mu_{N}$ on the WSM side), while it exhibits the typical linear behaviour like in 
metals for higher doping. We find that the Lorentz number shows  interesting features below the superconducting gap. This includes the violation of the Wiedemann-Franz law for $-\Delta_0<\mu_N<
\Delta_0$ and agreement (metal like) for higher doping concentrations ($\lvert \mu_{N}\rvert \gg \Delta_{0}$). Further, we have studied the thermopower/Seebeck co-efficient and the figure of merit. 
Surprisingly, the figure of merit shows a sharp increase for transport near the Weyl point, while it stays close to unity away from it. In addition, we have also investigated the effect of a thin insulating 
barrier on the thermal conductance and thermopower. Importantly, we find larger oscillations (with period $\pi$) in the thermal conductance with the variation of barrier strength for larger doping and 
negligible oscillation amplitudes for smaller doping concentrations. Moreover, very close to the Weyl point, the thermopower is shown to change sign from positive to negative, as we increase the 
barrier strength $\chi$.

As far as a practical realization of our setup is concerned, WSM-WSC heretrostructures may be possible to fabricate with appropriate materials; although, the material realization of WSC is still not known to the best of our knowledge. Within our setup, the  
chemical potentials $\mu_{N}$ and $\mu_{S}$ can be tuned via two additional external gate voltages. These gate voltages can modulate the Fermi energy (doping level) of the normal and 
superconducting sides of the junction. In particular, one gate voltage can tune $\mu_{N}$ over the entire sample. The other chemical potential $\mu_{S}=\mu_{N}+U_{0}$ can be independently 
modulated by  tuning $U_{0}$ by another gate voltage. For a typical value of $\Delta_{0}\sim 1.3~\rm meV$, it may be possible to achieve a high value of $zT\sim 9 -10$ at low temperatures 
($T\sim 1-2~\rm {K}$) and chemical potentials of $\mu_{N}\sim 0.15~\rm meV$.

\acknowledgments{}
\noindent
RS acknowledges support from the European Union's Horizon 2020 Research and Innovation Programme under Grant Agreement No. 766714/HiTIMe. PC acknowledges Debabrata  
Sinha and Arnob Kumar Ghosh for useful discussions.


\appendix

    \section{Derivation of Eq.~(\ref{current}) \ie the total current}
    \label{NS_tempbias}

\begin{figure}[!ht]
        \begin{center}
        \includegraphics[width=3.5in,height=0.8in]{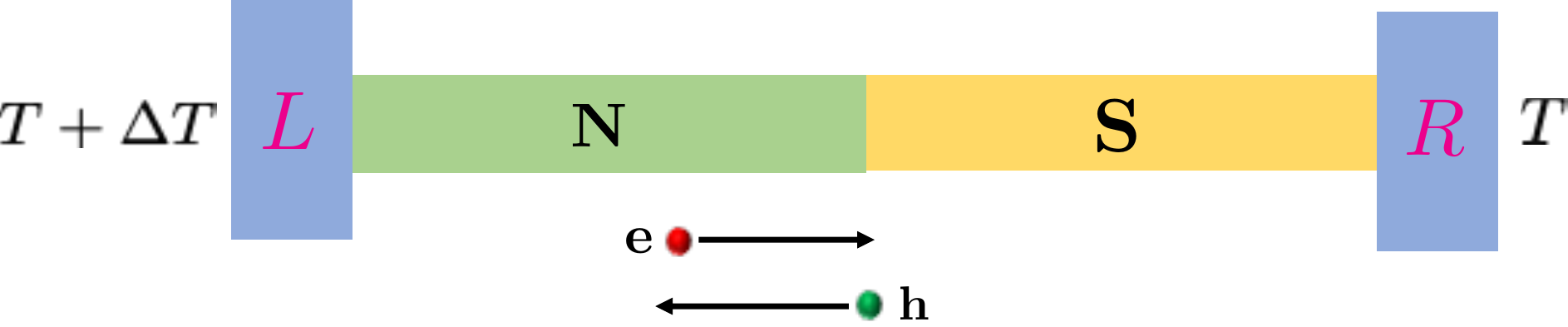}
        \end{center}
        \caption{\hspace*{-0.1cm}(Color online) Cartoon of a NS junction that is attached to left ($L$) and right ($R$) reservoirs which are kept at temperatures $T+\Delta T$ and $T$ respectively. 
        The red and green dots represent the flow of electrons and holes respectively due to the applied thermal bias.}
        \label{fig_app}
    \end{figure}

             We follow Ref.~\cite{BTK} to arrive at Eq.~(\ref{current}) (see in main text) for one-dimensional transport. This derivation is potentially important for understanding thermoelectric 
             coefficients $L_{12}$. 
             This physical quantity plays a significant role in determining the thermoelectric properties of any device, namely thermopower or Seeback coefficient. Thermoelectric current (determined by 
             thermoelectric coefficient) is conceptually a net charge transfer due to temperature bias. We schematically show the normal-superconductor (NS) junction in Fig.~\ref{fig_app} where we assume  
             that the left and right reservoirs are kept at bias voltages $E-e\Delta V$ and $E$ respectively as well as at temperatures $T+\Delta T$ and $T$ respectively. The flow of electrons from 
             the higher to the lower temperature region would cause a flow of electric current from left to right, as depicted by the electron flow indicated by a red dot in Fig.~\ref{fig_app}. The flow direction 
             of the Andreev reflected hole is denoted by a green dot. 
             \vspace{+0.2cm}

    The total current through the NS junction can be read from Ref.~\cite{BTK} as
    \begin{eqnarray}
    I=\frac{2e}{h} \int_{-\infty}^\infty \Big[ f_\rightarrow (E) - f_\leftarrow (E) \Big] dE \ ,
    \end{eqnarray}
    where, $f_\rightarrow (E)$ and $f_\leftarrow (E)$ are the distribution functions of particles coming from N and S side of the junction respectively.

       The distribution functions for finite voltage and temperature bias are given by
    \begin{align}
    f_\rightarrow (E) &= f_0 (E-e\Delta V,T+\Delta T), \nonumber\\
    f_\leftarrow (E) &= A(E) \Big[1-f_\rightarrow(-E,T-\Delta T)\Big] \nonumber \\ 
    &\hspace*{0.5cm}+B(E) f_\rightarrow(E,T+\Delta T) \nonumber \\ 
    &\hspace*{0.5cm}+ \left[C(E)+D(E)\right] f_0(E,T) \nonumber \\
    &= A(E) \, f_0(E+e\Delta V,T-\Delta T) \nonumber\\
    &\hspace*{0.5cm}+B(E) f_0(E-e\Delta V,T+\Delta T) \nonumber\\
    &\hspace*{0.5cm}+ \left[1-A(E)-B(E)\right] f_0(E,T) \ , 
    \end{align}
    where, we use $1-f_\rightarrow (-E,T-\Delta T)=f_0 (E+e\Delta V, T-\Delta T)$ and $A(E)+B(E)+C(E)+D(E)=1$ to arrive at the last expression. 
    We then Taylor expand the distribution functions for small voltage and temperature biases (linear response regime) as,
    \begin{align}
    f_0(E-e\Delta V,T+\Delta T) &= f_0(E,T)-e\Delta V \frac{\partial f_0}{\partial E}+\Delta T \frac{\partial f_0}{\partial T}\ , \nonumber \\
    f_0(E+e\Delta V,T-\Delta T) &= f_0(E,T)+e\Delta V \frac{\partial f_0}{\partial E}-\Delta T \frac{\partial f_0}{\partial T}\ .
    \label{Tailor_expand}
    \end{align}
    Using Eq.~(\ref{Tailor_expand}), we arrive at the current expression for one-dimensional transport as
    \begin{eqnarray}
    I=\frac{2e}{h} \int_{-\infty}^\infty \Big[-e\Delta V \frac{\partial f_0}{\partial E}+\Delta T \frac{\partial f_0}{\partial T} \Big](1+A-B) \, dE\ . \nonumber
    \end{eqnarray}
    
    Therefore, from the Onsager relation (see Eq.~(\ref{Onsager})), one can arrive at the expression for $L_{12}$ as~\cite{Bagwell1993}
    \begin{align}
    L_{12}=-\frac{2 e}{h} \int_{-\infty}^{\infty} \left(-\frac{\partial f_0}{\partial T}\right) (1+A(E)-B(E)) \,dE \ .
    \end{align}

    \section{Wiedemann-Franz law for bulk Weyl semimetal close to the Weyl point}
    \label{WF_law2}
    In this appendix, we provide an analytical derivation of WF-law for the three-dimensional (3D) bulk WSM near the Weyl point. The idea is to compare this with the Lorentz number for WSM-WSC 
    junction close to critical temperature ($T\sim T_{c}$) and $\mu_{N}\sim 0$ which is discussed in the main text and shown in Fig.~\ref{WFlawplot}. 
    We note that following the Blonder-Tinkham-Klapwijk   
    approach~\cite{BTK}, we numerically find the Lorentz number to be 4.1958 
    for $\mu_N=0$. Here, we follow a semiclassical approach which is adopted in a recent study of WF law in case of bulk graphene~\cite{AdamWFLAW2021}.
    
    The general expression of thermal conductivity $\kappa$ relating to the specific heat $C$ of the system is given by 
    \begin{eqnarray}
    \kappa=\frac{1}{d}\,C\,v_{F}\,\langle l\rangle\ , 
    \label{thcond}
    \end{eqnarray}
    where $d$, $v_F$ and $\langle l\rangle$ are the dimension, Fermi velocity and average mean free path respectively for the corresponding system.
    
    We compute the specific heat and the mean free path of a bulk 3D WSM considering the low energy effective linear spectrum $\epsilon_{\textcolor{blue}{p}}=\hbar v_{F} \lvert \textcolor{blue}{p} \rvert$. 
    Using the density of states close to the Weyl point as $D(E)=E^2/2\pi^2 (\hbar v_F)^3$, the specific heat of the system in terms of the internal energy is
    \begin{eqnarray}
    C=\frac{dU}{dT}=\frac{d}{dT} \left[\int_0^\infty dE\, D(E)\,f(E)\,E\right]\ ,
    \end{eqnarray}
    where $f(E)$ is the Fermi distribution function. After a few steps of algebra we arrive at the expression for $C$ in a WSM given by
    \begin{eqnarray}
    C=\frac{4}{\pi^{2}}\,\frac{k_{B}^{4}T^{3}}{(\hbar v_{F})^{3}}\,(5.6822)\ .
    \label{sh_app}
    \end{eqnarray}
    
    To calculate the mean free path, we use the Drude formula which relates the electrical conductivity to the mean free path of electrons in the system as
    \begin{eqnarray}
    G=\frac{n\,e^2\,l}{m^*\,v_F}\ .
    \end{eqnarray}
    where $m^*$ and $n$ denote the effective mass and number density of electrons respectively. 
    
    Using the number density $n=k_F^3/6\pi^2$ and considering only the transport along $z$ direction, we find the mean free path as
    \begin{eqnarray}
    l=G\, \frac{6\,\pi^2\,\hbar^3\, v_F^2}{\,E^2\,e^2}\ ,
    \end{eqnarray}

             Here, effective mass of the system in general is defined as
    \begin{eqnarray}
    m^{*}_{ij}=\frac{\hbar^{2}}{\left(\frac{\partial^{2}E}{\partial \textcolor{blue}{p}_{i}\partial \textcolor{blue}{p}_{j}}\right)} \ ,
    \end{eqnarray}
    For a spherical Fermi surface, off-diagonal components of the effective mass tensor become zero.  Therefore, the remaining terms are $m^{*}_{xx}=m^{*}_{yy}=m^{*}_{zz}
    =\frac{\hbar \textcolor{blue}{p}_{F}}{v_{F}}$. 
    For transport along $z$ direction, we have $m^{*}_{zz}=m^{*}=\frac{\hbar \textcolor{blue}{p}_{F}}{v_{F}}$.
    
    Hence, the average mean free path can be found as
    \vspace{-0.15cm}
    \begin{eqnarray}
         \langle l\rangle=G\, \frac{6\,\pi^2\,\hbar^3\, v_F^2}{\,e^2} \left< \frac{1}{E^2} \right>\ ,
    \end{eqnarray}
    Computing the average value of $1/E^2$ statistically we find the analytical expression of average mean free path as
    \begin{eqnarray}
              \langle l\rangle=G \frac{6\,\pi^2\,\hbar^3\, v_F^2}{\,e^2} \left< \frac{0.384423}{(k_B\,T^2)} \right>\ .
              \label{mfp_app}
    \end{eqnarray}
    
    Using the specific heat in Eq.~(\ref{sh_app}) and the average mean free path in Eq.~(\ref{mfp_app}) one can compute the thermal conductance employing Eq.~(\ref{thcond}). 
    Finally, substituting this quantity in Eq.~(\ref{WFlaw}) we arrive at the Lorentz number 
    \begin{eqnarray}
    L=\frac{\kappa}{G\,T}=5.3171 \times L_0 \ ,
    \end{eqnarray}
         where, $L_{0}=\frac{\pi^{2}}{3}(\frac{k_{B}}{e})^{2}$.

\bibliography{lib}{}

\begin{thebibliography}{60}%
\makeatletter
\providecommand \@ifxundefined [1]{%
 \@ifx{#1\undefined}
}%
\providecommand \@ifnum [1]{%
 \ifnum #1\expandafter \@firstoftwo
 \else \expandafter \@secondoftwo
 \fi
}%
\providecommand \@ifx [1]{%
 \ifx #1\expandafter \@firstoftwo
 \else \expandafter \@secondoftwo
 \fi
}%
\providecommand \natexlab [1]{#1}%
\providecommand \enquote  [1]{``#1''}%
\providecommand \bibnamefont  [1]{#1}%
\providecommand \bibfnamefont [1]{#1}%
\providecommand \citenamefont [1]{#1}%
\providecommand \href@noop [0]{\@secondoftwo}%
\providecommand \href [0]{\begingroup \@sanitize@url \@href}%
\providecommand \@href[1]{\@@startlink{#1}\@@href}%
\providecommand \@@href[1]{\endgroup#1\@@endlink}%
\providecommand \@sanitize@url [0]{\catcode `\\12\catcode `\$12\catcode
  `\&12\catcode `\#12\catcode `\^12\catcode `\_12\catcode `\%12\relax}%
\providecommand \@@startlink[1]{}%
\providecommand \@@endlink[0]{}%
\providecommand \url  [0]{\begingroup\@sanitize@url \@url }%
\providecommand \@url [1]{\endgroup\@href {#1}{\urlprefix }}%
\providecommand \urlprefix  [0]{URL }%
\providecommand \Eprint [0]{\href }%
\providecommand \doibase [0]{http://dx.doi.org/}%
\providecommand \selectlanguage [0]{\@gobble}%
\providecommand \bibinfo  [0]{\@secondoftwo}%
\providecommand \bibfield  [0]{\@secondoftwo}%
\providecommand \translation [1]{[#1]}%
\providecommand \BibitemOpen [0]{}%
\providecommand \bibitemStop [0]{}%
\providecommand \bibitemNoStop [0]{.\EOS\space}%
\providecommand \EOS [0]{\spacefactor3000\relax}%
\providecommand \BibitemShut  [1]{\csname bibitem#1\endcsname}%
\let\auto@bib@innerbib\@empty
\bibitem [{\citenamefont {Adler}(1969)}]{Adler}%
  \BibitemOpen
  \bibfield  {author} {\bibinfo {author} {\bibfnamefont {Stephen~L.}\
  \bibnamefont {Adler}},\ }\bibfield  {title} {\enquote {\bibinfo {title}
  {Axial-vector vertex in spinor electrodynamics},}\ }\href {\doibase
  10.1103/PhysRev.177.2426} {\bibfield  {journal} {\bibinfo  {journal} {Phys.
  Rev.}\ }\textbf {\bibinfo {volume} {177}},\ \bibinfo {pages} {2426--2438}
  (\bibinfo {year} {1969})}\BibitemShut {NoStop}%
\bibitem [{\citenamefont {Bell}\ and\ \citenamefont
  {Jackiw}(1969)}]{BellJackiw}%
  \BibitemOpen
  \bibfield  {author} {\bibinfo {author} {\bibfnamefont {J.~S.}\ \bibnamefont
  {Bell}}\ and\ \bibinfo {author} {\bibfnamefont {R.~A}\ \bibnamefont
  {Jackiw}},\ }\bibfield  {title} {\enquote {\bibinfo {title} {Pcac puzzle:
  $\pi^0\rightarrow \gamma \gamma$ in the $\sigma$-model.}}\ }\href {\doibase
  10.1007/BF02823296} {\bibfield  {journal} {\bibinfo  {journal} {Nuovo Cimento
  A (1965-1970)}\ }\textbf {\bibinfo {volume} {60}},\ \bibinfo {pages} {47--61}
  (\bibinfo {year} {1969})}\BibitemShut {NoStop}%
\bibitem [{\citenamefont {Fukushima}\ \emph {et~al.}(2008)\citenamefont
  {Fukushima}, \citenamefont {Kharzeev},\ and\ \citenamefont
  {Warringa}}]{PhysRevD.78.074033}%
  \BibitemOpen
  \bibfield  {author} {\bibinfo {author} {\bibfnamefont {Kenji}\ \bibnamefont
  {Fukushima}}, \bibinfo {author} {\bibfnamefont {Dmitri~E.}\ \bibnamefont
  {Kharzeev}}, \ and\ \bibinfo {author} {\bibfnamefont {Harmen~J.}\
  \bibnamefont {Warringa}},\ }\bibfield  {title} {\enquote {\bibinfo {title}
  {Chiral magnetic effect},}\ }\href {\doibase 10.1103/PhysRevD.78.074033}
  {\bibfield  {journal} {\bibinfo  {journal} {Phys. Rev. D}\ }\textbf {\bibinfo
  {volume} {78}},\ \bibinfo {pages} {074033} (\bibinfo {year}
  {2008})}\BibitemShut {NoStop}%
\bibitem [{\citenamefont {Armitage}\ \emph {et~al.}(2018)\citenamefont
  {Armitage}, \citenamefont {Mele},\ and\ \citenamefont
  {Vishwanath}}]{Armitage2018}%
  \BibitemOpen
  \bibfield  {author} {\bibinfo {author} {\bibfnamefont {N.~P.}\ \bibnamefont
  {Armitage}}, \bibinfo {author} {\bibfnamefont {E.~J.}\ \bibnamefont {Mele}},
  \ and\ \bibinfo {author} {\bibfnamefont {Ashvin}\ \bibnamefont
  {Vishwanath}},\ }\bibfield  {title} {\enquote {\bibinfo {title} {Weyl and
  dirac semimetals in three-dimensional solids},}\ }\href {\doibase
  10.1103/RevModPhys.90.015001} {\bibfield  {journal} {\bibinfo  {journal}
  {Rev. Mod. Phys.}\ }\textbf {\bibinfo {volume} {90}},\ \bibinfo {pages}
  {015001} (\bibinfo {year} {2018})}\BibitemShut {NoStop}%
\bibitem [{\citenamefont {Huang}\ \emph {et~al.}(2015)\citenamefont {Huang},
  \citenamefont {Xu}, \citenamefont {Belopolski}, \citenamefont {Lee},
  \citenamefont {Chang}, \citenamefont {Wang}, \citenamefont {Alidoust},
  \citenamefont {Bian}, \citenamefont {Neupane}, \citenamefont {Zhang},
  \citenamefont {Jia}, \citenamefont {Bansil}, \citenamefont {Lin},\ and\
  \citenamefont {Hasan}}]{Hasan2015}%
  \BibitemOpen
  \bibfield  {author} {\bibinfo {author} {\bibfnamefont {Shin-Ming}\
  \bibnamefont {Huang}}, \bibinfo {author} {\bibfnamefont {Su-Yang}\
  \bibnamefont {Xu}}, \bibinfo {author} {\bibfnamefont {Ilya}\ \bibnamefont
  {Belopolski}}, \bibinfo {author} {\bibfnamefont {Chi-Cheng}\ \bibnamefont
  {Lee}}, \bibinfo {author} {\bibfnamefont {Guoqing}\ \bibnamefont {Chang}},
  \bibinfo {author} {\bibfnamefont {BaoKai}\ \bibnamefont {Wang}}, \bibinfo
  {author} {\bibfnamefont {Nasser}\ \bibnamefont {Alidoust}}, \bibinfo {author}
  {\bibfnamefont {Guang}\ \bibnamefont {Bian}}, \bibinfo {author}
  {\bibfnamefont {Madhab}\ \bibnamefont {Neupane}}, \bibinfo {author}
  {\bibfnamefont {Chenglong}\ \bibnamefont {Zhang}}, \bibinfo {author}
  {\bibfnamefont {Shuang}\ \bibnamefont {Jia}}, \bibinfo {author}
  {\bibfnamefont {Arun}\ \bibnamefont {Bansil}}, \bibinfo {author}
  {\bibfnamefont {Hsin}\ \bibnamefont {Lin}}, \ and\ \bibinfo {author}
  {\bibfnamefont {M.~Zahid}\ \bibnamefont {Hasan}},\ }\bibfield  {title}
  {\enquote {\bibinfo {title} {A {W}eyl {F}ermion semimetal with surface fermi
  arcs in the transition metal monopnictide {T}a{A}s class},}\ }\href
  {https://doi.org/10.1038/ncomms8373} {\bibfield  {journal} {\bibinfo
  {journal} {Nature Communications}\ }\textbf {\bibinfo {volume} {6}},\
  \bibinfo {pages} {7373} (\bibinfo {year} {2015})}\BibitemShut {NoStop}%
\bibitem [{\citenamefont {Xu}\ \emph {et~al.}(2015)\citenamefont {Xu},
  \citenamefont {Belopolski}, \citenamefont {Sanchez}, \citenamefont {Zhang},
  \citenamefont {Chang}, \citenamefont {Guo}, \citenamefont {Bian},
  \citenamefont {Yuan}, \citenamefont {Lu}, \citenamefont {Chang},
  \citenamefont {Shibayev}, \citenamefont {Prokopovych}, \citenamefont
  {Alidoust}, \citenamefont {Zheng}, \citenamefont {Lee}, \citenamefont
  {Huang}, \citenamefont {Sankar}, \citenamefont {Chou}, \citenamefont {Hsu},
  \citenamefont {Jeng}, \citenamefont {Bansil}, \citenamefont {Neupert},
  \citenamefont {Strocov}, \citenamefont {Lin}, \citenamefont {Jia},\ and\
  \citenamefont {Hasan}}]{2Hasan2015}%
  \BibitemOpen
  \bibfield  {author} {\bibinfo {author} {\bibfnamefont {S.~Y.}\ \bibnamefont
  {Xu}}, \bibinfo {author} {\bibfnamefont {I.}~\bibnamefont {Belopolski}},
  \bibinfo {author} {\bibfnamefont {D.~S.}\ \bibnamefont {Sanchez}}, \bibinfo
  {author} {\bibfnamefont {C.~L.}\ \bibnamefont {Zhang}}, \bibinfo {author}
  {\bibfnamefont {G.~Q.}\ \bibnamefont {Chang}}, \bibinfo {author}
  {\bibfnamefont {C.}~\bibnamefont {Guo}}, \bibinfo {author} {\bibfnamefont
  {G.}~\bibnamefont {Bian}}, \bibinfo {author} {\bibfnamefont {Z.~J.}\
  \bibnamefont {Yuan}}, \bibinfo {author} {\bibfnamefont {H.}~\bibnamefont
  {Lu}}, \bibinfo {author} {\bibfnamefont {T.~R.}\ \bibnamefont {Chang}},
  \bibinfo {author} {\bibfnamefont {P.~P.}\ \bibnamefont {Shibayev}}, \bibinfo
  {author} {\bibfnamefont {M.~L.}\ \bibnamefont {Prokopovych}}, \bibinfo
  {author} {\bibfnamefont {N.}~\bibnamefont {Alidoust}}, \bibinfo {author}
  {\bibfnamefont {H.}~\bibnamefont {Zheng}}, \bibinfo {author} {\bibfnamefont
  {C.~C.}\ \bibnamefont {Lee}}, \bibinfo {author} {\bibfnamefont {S.~M.}\
  \bibnamefont {Huang}}, \bibinfo {author} {\bibfnamefont {R.}~\bibnamefont
  {Sankar}}, \bibinfo {author} {\bibfnamefont {F.~C.}\ \bibnamefont {Chou}},
  \bibinfo {author} {\bibfnamefont {C.~H.}\ \bibnamefont {Hsu}}, \bibinfo
  {author} {\bibfnamefont {H.~T.}\ \bibnamefont {Jeng}}, \bibinfo {author}
  {\bibfnamefont {A.}~\bibnamefont {Bansil}}, \bibinfo {author} {\bibfnamefont
  {T.}~\bibnamefont {Neupert}}, \bibinfo {author} {\bibfnamefont {V.~N.}\
  \bibnamefont {Strocov}}, \bibinfo {author} {\bibfnamefont {H.}~\bibnamefont
  {Lin}}, \bibinfo {author} {\bibfnamefont {S.}~\bibnamefont {Jia}}, \ and\
  \bibinfo {author} {\bibfnamefont {M.~Z.}\ \bibnamefont {Hasan}},\ }\bibfield
  {title} {\enquote {\bibinfo {title} {Experimental discovery of a topological
  {W}ey semimetal state in {T}a{P}},}\ }\href {10.1126/sciadv.1501092}
  {\bibfield  {journal} {\bibinfo  {journal} {Science Advances}\ }\textbf
  {\bibinfo {volume} {1}},\ \bibinfo {pages} {e1501092} (\bibinfo {year}
  {2015})}\BibitemShut {NoStop}%
\bibitem [{\citenamefont {Vazifeh}\ and\ \citenamefont
  {Franz}(2013)}]{Franz2013}%
  \BibitemOpen
  \bibfield  {author} {\bibinfo {author} {\bibfnamefont {M.~M.}\ \bibnamefont
  {Vazifeh}}\ and\ \bibinfo {author} {\bibfnamefont {M.}~\bibnamefont
  {Franz}},\ }\bibfield  {title} {\enquote {\bibinfo {title} {Electromagnetic
  response of weyl semimetals},}\ }\href {\doibase
  10.1103/PhysRevLett.111.027201} {\bibfield  {journal} {\bibinfo  {journal}
  {Phys. Rev. Lett.}\ }\textbf {\bibinfo {volume} {111}},\ \bibinfo {pages}
  {027201} (\bibinfo {year} {2013})}\BibitemShut {NoStop}%
\bibitem [{\citenamefont {Zyuzin}\ and\ \citenamefont
  {Burkov}(2012)}]{Burkov2012}%
  \BibitemOpen
  \bibfield  {author} {\bibinfo {author} {\bibfnamefont {A.~A.}\ \bibnamefont
  {Zyuzin}}\ and\ \bibinfo {author} {\bibfnamefont {A.~A.}\ \bibnamefont
  {Burkov}},\ }\bibfield  {title} {\enquote {\bibinfo {title} {Topological
  response in weyl semimetals and the chiral anomaly},}\ }\href {\doibase
  10.1103/PhysRevB.86.115133} {\bibfield  {journal} {\bibinfo  {journal} {Phys.
  Rev. B}\ }\textbf {\bibinfo {volume} {86}},\ \bibinfo {pages} {115133}
  (\bibinfo {year} {2012})}\BibitemShut {NoStop}%
\bibitem [{\citenamefont {Lv}\ \emph {et~al.}(2015{\natexlab{a}})\citenamefont
  {Lv}, \citenamefont {Weng}, \citenamefont {Fu}, \citenamefont {Wang},
  \citenamefont {Miao}, \citenamefont {Ma}, \citenamefont {Richard},
  \citenamefont {Huang}, \citenamefont {Zhao}, \citenamefont {Chen},
  \citenamefont {Fang}, \citenamefont {Dai}, \citenamefont {Qian},\ and\
  \citenamefont {Ding}}]{Ding2005}%
  \BibitemOpen
  \bibfield  {author} {\bibinfo {author} {\bibfnamefont {B.~Q.}\ \bibnamefont
  {Lv}}, \bibinfo {author} {\bibfnamefont {H.~M.}\ \bibnamefont {Weng}},
  \bibinfo {author} {\bibfnamefont {B.~B.}\ \bibnamefont {Fu}}, \bibinfo
  {author} {\bibfnamefont {X.~P.}\ \bibnamefont {Wang}}, \bibinfo {author}
  {\bibfnamefont {H.}~\bibnamefont {Miao}}, \bibinfo {author} {\bibfnamefont
  {J.}~\bibnamefont {Ma}}, \bibinfo {author} {\bibfnamefont {P.}~\bibnamefont
  {Richard}}, \bibinfo {author} {\bibfnamefont {X.~C.}\ \bibnamefont {Huang}},
  \bibinfo {author} {\bibfnamefont {L.~X.}\ \bibnamefont {Zhao}}, \bibinfo
  {author} {\bibfnamefont {G.~F.}\ \bibnamefont {Chen}}, \bibinfo {author}
  {\bibfnamefont {Z.}~\bibnamefont {Fang}}, \bibinfo {author} {\bibfnamefont
  {X.}~\bibnamefont {Dai}}, \bibinfo {author} {\bibfnamefont {T.}~\bibnamefont
  {Qian}}, \ and\ \bibinfo {author} {\bibfnamefont {H.}~\bibnamefont {Ding}},\
  }\bibfield  {title} {\enquote {\bibinfo {title} {Experimental discovery of
  weyl semimetal taas},}\ }\href {\doibase 10.1103/PhysRevX.5.031013}
  {\bibfield  {journal} {\bibinfo  {journal} {Phys. Rev. X}\ }\textbf {\bibinfo
  {volume} {5}},\ \bibinfo {pages} {031013} (\bibinfo {year}
  {2015}{\natexlab{a}})}\BibitemShut {NoStop}%
\bibitem [{\citenamefont {Burkov}\ \emph {et~al.}(2011)\citenamefont {Burkov},
  \citenamefont {Hook},\ and\ \citenamefont {Balents}}]{Burkov2011}%
  \BibitemOpen
  \bibfield  {author} {\bibinfo {author} {\bibfnamefont {A.~A.}\ \bibnamefont
  {Burkov}}, \bibinfo {author} {\bibfnamefont {M.~D.}\ \bibnamefont {Hook}}, \
  and\ \bibinfo {author} {\bibfnamefont {Leon}\ \bibnamefont {Balents}},\
  }\bibfield  {title} {\enquote {\bibinfo {title} {Topological nodal
  semimetals},}\ }\href {\doibase 10.1103/PhysRevB.84.235126} {\bibfield
  {journal} {\bibinfo  {journal} {Phys. Rev. B}\ }\textbf {\bibinfo {volume}
  {84}},\ \bibinfo {pages} {235126} (\bibinfo {year} {2011})}\BibitemShut
  {NoStop}%
\bibitem [{\citenamefont {Yan}\ and\ \citenamefont
  {Felser}(2017)}]{Claudia2017}%
  \BibitemOpen
  \bibfield  {author} {\bibinfo {author} {\bibfnamefont {Binghai}\ \bibnamefont
  {Yan}}\ and\ \bibinfo {author} {\bibfnamefont {Claudia}\ \bibnamefont
  {Felser}},\ }\bibfield  {title} {\enquote {\bibinfo {title} {Topological
  materials: {W}eyl {S}emimetals},}\ }\href
  {https://doi.org/10.1146/annurev-conmatphys-031016-025458} {\bibfield
  {journal} {\bibinfo  {journal} {Annual Review of Condensed Matter Physics}\
  }\textbf {\bibinfo {volume} {8}},\ \bibinfo {pages} {337--354} (\bibinfo
  {year} {2017})}\BibitemShut {NoStop}%
\bibitem [{\citenamefont {Zhou}\ \emph {et~al.}(2019)\citenamefont {Zhou},
  \citenamefont {Zhao}, \citenamefont {Zeng}, \citenamefont {Chen},\ and\
  \citenamefont {Geng}}]{Yun2019}%
  \BibitemOpen
  \bibfield  {author} {\bibinfo {author} {\bibfnamefont {Yu}~\bibnamefont
  {Zhou}}, \bibinfo {author} {\bibfnamefont {Ying-Qin}\ \bibnamefont {Zhao}},
  \bibinfo {author} {\bibfnamefont {Zhao-Yi}\ \bibnamefont {Zeng}}, \bibinfo
  {author} {\bibfnamefont {Xiang-Rong}\ \bibnamefont {Chen}}, \ and\ \bibinfo
  {author} {\bibfnamefont {Hua-Yun}\ \bibnamefont {Geng}},\ }\bibfield  {title}
  {\enquote {\bibinfo {title} {Anisotropic thermoelectric properties of {W}eyl
  semimetal {N}b{X} ({X} = {P} and {A}s): a potential thermoelectric
  material},}\ }\href {\doibase 10.1088/1361-648x/aa7a3b} {\bibfield  {journal}
  {\bibinfo  {journal} {Phys. Chem. Chem. Phys.}\ }\textbf {\bibinfo {volume}
  {21}},\ \bibinfo {pages} {15167--15176} (\bibinfo {year} {2019})}\BibitemShut
  {NoStop}%
\bibitem [{\citenamefont {Chang}\ \emph {et~al.}(2016)\citenamefont {Chang},
  \citenamefont {Liu}, \citenamefont {Rao}, \citenamefont {Wang}, \citenamefont
  {Sunac},\ and\ \citenamefont {Jia}}]{Yu2016}%
  \BibitemOpen
  \bibfield  {author} {\bibinfo {author} {\bibfnamefont {D.}~\bibnamefont
  {Chang}}, \bibinfo {author} {\bibfnamefont {Y.}~\bibnamefont {Liu}}, \bibinfo
  {author} {\bibfnamefont {F.}~\bibnamefont {Rao}}, \bibinfo {author}
  {\bibfnamefont {F.}~\bibnamefont {Wang}}, \bibinfo {author} {\bibfnamefont
  {Q.}~\bibnamefont {Sunac}}, \ and\ \bibinfo {author} {\bibfnamefont
  {Y.}~\bibnamefont {Jia}},\ }\bibfield  {title} {\enquote {\bibinfo {title}
  {Phonon and thermal expansion properties in {W}eyl semimetals {MX} ({M} =
  {N}b{,} {T}a; {X} = {P}{,} {A}s): ab initio studies},}\ }\href
  {https://doi.org/10.1039/C6CP02018F} {\bibfield  {journal} {\bibinfo
  {journal} {Phys. Chem. Phys.}\ }\textbf {\bibinfo {volume} {18}},\ \bibinfo
  {pages} {14503--14508} (\bibinfo {year} {2016})}\BibitemShut {NoStop}%
\bibitem [{\citenamefont {Sun}\ \emph {et~al.}(2015)\citenamefont {Sun},
  \citenamefont {Wu},\ and\ \citenamefont {Yan}}]{Yan2015}%
  \BibitemOpen
  \bibfield  {author} {\bibinfo {author} {\bibfnamefont {Yan}\ \bibnamefont
  {Sun}}, \bibinfo {author} {\bibfnamefont {Shu-Chun}\ \bibnamefont {Wu}}, \
  and\ \bibinfo {author} {\bibfnamefont {Binghai}\ \bibnamefont {Yan}},\
  }\bibfield  {title} {\enquote {\bibinfo {title} {Topological surface states
  and fermi arcs of the noncentrosymmetric weyl semimetals taas, tap, nbas, and
  nbp},}\ }\href {\doibase 10.1103/PhysRevB.92.115428} {\bibfield  {journal}
  {\bibinfo  {journal} {Phys. Rev. B}\ }\textbf {\bibinfo {volume} {92}},\
  \bibinfo {pages} {115428} (\bibinfo {year} {2015})}\BibitemShut {NoStop}%
\bibitem [{\citenamefont {Lv}\ \emph {et~al.}(2015{\natexlab{b}})\citenamefont
  {Lv}, \citenamefont {Xu}, \citenamefont {Weng}, \citenamefont {Ma},
  \citenamefont {Richard}, \citenamefont {Huang}, \citenamefont {Zhao},
  \citenamefont {Chen}, \citenamefont {Matt}, \citenamefont {Bisti},
  \citenamefont {Strocov}, \citenamefont {Mesot}, \citenamefont {Fang},
  \citenamefont {Dai}, \citenamefont {Qian}, \citenamefont {Shi},\ and\
  \citenamefont {Ding}}]{Ding2015}%
  \BibitemOpen
  \bibfield  {author} {\bibinfo {author} {\bibfnamefont {B.~Q.}\ \bibnamefont
  {Lv}}, \bibinfo {author} {\bibfnamefont {N.}~\bibnamefont {Xu}}, \bibinfo
  {author} {\bibfnamefont {H.~M.}\ \bibnamefont {Weng}}, \bibinfo {author}
  {\bibfnamefont {J.~Z.}\ \bibnamefont {Ma}}, \bibinfo {author} {\bibfnamefont
  {P.}~\bibnamefont {Richard}}, \bibinfo {author} {\bibfnamefont {X.~C.}\
  \bibnamefont {Huang}}, \bibinfo {author} {\bibfnamefont {L.~X.}\ \bibnamefont
  {Zhao}}, \bibinfo {author} {\bibfnamefont {G.~F.}\ \bibnamefont {Chen}},
  \bibinfo {author} {\bibfnamefont {C.~E.}\ \bibnamefont {Matt}}, \bibinfo
  {author} {\bibfnamefont {F.}~\bibnamefont {Bisti}}, \bibinfo {author}
  {\bibfnamefont {V.~N.}\ \bibnamefont {Strocov}}, \bibinfo {author}
  {\bibfnamefont {J.}~\bibnamefont {Mesot}}, \bibinfo {author} {\bibfnamefont
  {Z.}~\bibnamefont {Fang}}, \bibinfo {author} {\bibfnamefont {X.}~\bibnamefont
  {Dai}}, \bibinfo {author} {\bibfnamefont {T.}~\bibnamefont {Qian}}, \bibinfo
  {author} {\bibfnamefont {M.}~\bibnamefont {Shi}}, \ and\ \bibinfo {author}
  {\bibfnamefont {H.}~\bibnamefont {Ding}},\ }\bibfield  {title} {\enquote
  {\bibinfo {title} {Observation of {W}eyl nodes in {T}a{A}s},}\ }\href
  {https://doi.org/10.1038/nphys3426} {\bibfield  {journal} {\bibinfo
  {journal} {Nature Physics}\ }\textbf {\bibinfo {volume} {11}},\ \bibinfo
  {pages} {724--727} (\bibinfo {year} {2015}{\natexlab{b}})}\BibitemShut
  {NoStop}%
\bibitem [{\citenamefont {Yang}\ \emph {et~al.}(2015)\citenamefont {Yang},
  \citenamefont {Liu}, \citenamefont {Sun}, \citenamefont {Peng}, \citenamefont
  {Yang}, \citenamefont {Zhang}, \citenamefont {Zhou}, \citenamefont {Zhang},
  \citenamefont {Guo}, \citenamefont {Rahn}, \citenamefont {Prabhakaran},
  \citenamefont {Hussain}, \citenamefont {Mo}, \citenamefont {Felser},
  \citenamefont {Yan},\ and\ \citenamefont {Chen}}]{Claudia2015}%
  \BibitemOpen
  \bibfield  {author} {\bibinfo {author} {\bibfnamefont {L.~X.}\ \bibnamefont
  {Yang}}, \bibinfo {author} {\bibfnamefont {Z.~K.}\ \bibnamefont {Liu}},
  \bibinfo {author} {\bibfnamefont {Y.}~\bibnamefont {Sun}}, \bibinfo {author}
  {\bibfnamefont {H.}~\bibnamefont {Peng}}, \bibinfo {author} {\bibfnamefont
  {H.~F.}\ \bibnamefont {Yang}}, \bibinfo {author} {\bibfnamefont
  {T.}~\bibnamefont {Zhang}}, \bibinfo {author} {\bibfnamefont
  {B.}~\bibnamefont {Zhou}}, \bibinfo {author} {\bibfnamefont {Y.}~\bibnamefont
  {Zhang}}, \bibinfo {author} {\bibfnamefont {Y.~F.}\ \bibnamefont {Guo}},
  \bibinfo {author} {\bibfnamefont {M.}~\bibnamefont {Rahn}}, \bibinfo {author}
  {\bibfnamefont {D.}~\bibnamefont {Prabhakaran}}, \bibinfo {author}
  {\bibfnamefont {Z.}~\bibnamefont {Hussain}}, \bibinfo {author} {\bibfnamefont
  {S.~K.}\ \bibnamefont {Mo}}, \bibinfo {author} {\bibfnamefont
  {C.}~\bibnamefont {Felser}}, \bibinfo {author} {\bibfnamefont
  {B.}~\bibnamefont {Yan}}, \ and\ \bibinfo {author} {\bibfnamefont {Y.~L.}\
  \bibnamefont {Chen}},\ }\bibfield  {title} {\enquote {\bibinfo {title} {Weyl
  semimetal phase in the non-centrosymmetric compound {T}a{A}s},}\ }\href
  {https://doi.org/10.1038/nphys3425} {\bibfield  {journal} {\bibinfo
  {journal} {Nature Physics}\ }\textbf {\bibinfo {volume} {11}},\ \bibinfo
  {pages} {728--732} (\bibinfo {year} {2015})}\BibitemShut {NoStop}%
\bibitem [{\citenamefont {Chandrasekhar}(2009)}]{Chandrasekhar_2009}%
  \BibitemOpen
  \bibfield  {author} {\bibinfo {author} {\bibfnamefont {Venkat}\ \bibnamefont
  {Chandrasekhar}},\ }\bibfield  {title} {\enquote {\bibinfo {title} {Thermal
  transport in superconductor/normal-metal structures},}\ }\href {\doibase
  10.1088/0953-2048/22/8/083001} {\bibfield  {journal} {\bibinfo  {journal}
  {Superconductor Science and Technology}\ }\textbf {\bibinfo {volume} {22}},\
  \bibinfo {pages} {083001} (\bibinfo {year} {2009})}\BibitemShut {NoStop}%
\bibitem [{\citenamefont {Machon}\ \emph {et~al.}(2014)\citenamefont {Machon},
  \citenamefont {Eschrig},\ and\ \citenamefont {Belzig}}]{Machon_2014}%
  \BibitemOpen
  \bibfield  {author} {\bibinfo {author} {\bibfnamefont {P}~\bibnamefont
  {Machon}}, \bibinfo {author} {\bibfnamefont {M}~\bibnamefont {Eschrig}}, \
  and\ \bibinfo {author} {\bibfnamefont {W}~\bibnamefont {Belzig}},\ }\bibfield
   {title} {\enquote {\bibinfo {title} {Giant thermoelectric effects in a
  proximity-coupled superconductor{\textendash}ferromagnet device},}\ }\href
  {\doibase 10.1088/1367-2630/16/7/073002} {\bibfield  {journal} {\bibinfo
  {journal} {New Journal of Physics}\ }\textbf {\bibinfo {volume} {16}},\
  \bibinfo {pages} {073002} (\bibinfo {year} {2014})}\BibitemShut {NoStop}%
\bibitem [{\citenamefont {Ozaeta}\ \emph {et~al.}(2014)\citenamefont {Ozaeta},
  \citenamefont {Virtanen}, \citenamefont {Bergeret},\ and\ \citenamefont
  {Heikkil\"a}}]{PhysRevLett.112.057001}%
  \BibitemOpen
  \bibfield  {author} {\bibinfo {author} {\bibfnamefont {A.}~\bibnamefont
  {Ozaeta}}, \bibinfo {author} {\bibfnamefont {P.}~\bibnamefont {Virtanen}},
  \bibinfo {author} {\bibfnamefont {F.~S.}\ \bibnamefont {Bergeret}}, \ and\
  \bibinfo {author} {\bibfnamefont {T.~T.}\ \bibnamefont {Heikkil\"a}},\
  }\bibfield  {title} {\enquote {\bibinfo {title} {Predicted very large
  thermoelectric effect in ferromagnet-superconductor junctions in the presence
  of a spin-splitting magnetic field},}\ }\href {\doibase
  10.1103/PhysRevLett.112.057001} {\bibfield  {journal} {\bibinfo  {journal}
  {Phys. Rev. Lett.}\ }\textbf {\bibinfo {volume} {112}},\ \bibinfo {pages}
  {057001} (\bibinfo {year} {2014})}\BibitemShut {NoStop}%
\bibitem [{\citenamefont {Kalenkov}\ \emph {et~al.}(2012)\citenamefont
  {Kalenkov}, \citenamefont {Zaikin},\ and\ \citenamefont
  {Kuzmin}}]{PhysRevLett.109.147004}%
  \BibitemOpen
  \bibfield  {author} {\bibinfo {author} {\bibfnamefont {Mikhail~S.}\
  \bibnamefont {Kalenkov}}, \bibinfo {author} {\bibfnamefont {Andrei~D.}\
  \bibnamefont {Zaikin}}, \ and\ \bibinfo {author} {\bibfnamefont {Leonid~S.}\
  \bibnamefont {Kuzmin}},\ }\bibfield  {title} {\enquote {\bibinfo {title}
  {Theory of a large thermoelectric effect in superconductors doped with
  magnetic impurities},}\ }\href {\doibase 10.1103/PhysRevLett.109.147004}
  {\bibfield  {journal} {\bibinfo  {journal} {Phys. Rev. Lett.}\ }\textbf
  {\bibinfo {volume} {109}},\ \bibinfo {pages} {147004} (\bibinfo {year}
  {2012})}\BibitemShut {NoStop}%
\bibitem [{\citenamefont {Machon}\ \emph {et~al.}(2013)\citenamefont {Machon},
  \citenamefont {Eschrig},\ and\ \citenamefont
  {Belzig}}]{PhysRevLett.110.047002}%
  \BibitemOpen
  \bibfield  {author} {\bibinfo {author} {\bibfnamefont {P.}~\bibnamefont
  {Machon}}, \bibinfo {author} {\bibfnamefont {M.}~\bibnamefont {Eschrig}}, \
  and\ \bibinfo {author} {\bibfnamefont {W.}~\bibnamefont {Belzig}},\
  }\bibfield  {title} {\enquote {\bibinfo {title} {Nonlocal thermoelectric
  effects and nonlocal onsager relations in a three-terminal proximity-coupled
  superconductor-ferromagnet device},}\ }\href {\doibase
  10.1103/PhysRevLett.110.047002} {\bibfield  {journal} {\bibinfo  {journal}
  {Phys. Rev. Lett.}\ }\textbf {\bibinfo {volume} {110}},\ \bibinfo {pages}
  {047002} (\bibinfo {year} {2013})}\BibitemShut {NoStop}%
\bibitem [{\citenamefont {Kolenda}\ \emph {et~al.}(2016)\citenamefont
  {Kolenda}, \citenamefont {Wolf},\ and\ \citenamefont
  {Beckmann}}]{PhysRevLett.116.097001}%
  \BibitemOpen
  \bibfield  {author} {\bibinfo {author} {\bibfnamefont {S.}~\bibnamefont
  {Kolenda}}, \bibinfo {author} {\bibfnamefont {M.~J.}\ \bibnamefont {Wolf}}, \
  and\ \bibinfo {author} {\bibfnamefont {D.}~\bibnamefont {Beckmann}},\
  }\bibfield  {title} {\enquote {\bibinfo {title} {Observation of
  thermoelectric currents in high-field superconductor-ferromagnet tunnel
  junctions},}\ }\href {\doibase 10.1103/PhysRevLett.116.097001} {\bibfield
  {journal} {\bibinfo  {journal} {Phys. Rev. Lett.}\ }\textbf {\bibinfo
  {volume} {116}},\ \bibinfo {pages} {097001} (\bibinfo {year}
  {2016})}\BibitemShut {NoStop}%
\bibitem [{\citenamefont {Dutta}\ \emph {et~al.}(2017)\citenamefont {Dutta},
  \citenamefont {Saha},\ and\ \citenamefont {Jayannavar}}]{Arijit2017}%
  \BibitemOpen
  \bibfield  {author} {\bibinfo {author} {\bibfnamefont {Paramita}\
  \bibnamefont {Dutta}}, \bibinfo {author} {\bibfnamefont {Arijit}\
  \bibnamefont {Saha}}, \ and\ \bibinfo {author} {\bibfnamefont {A.~M.}\
  \bibnamefont {Jayannavar}},\ }\bibfield  {title} {\enquote {\bibinfo {title}
  {Thermoelectric properties of a ferromagnet-superconductor hybrid junction:
  Role of interfacial rashba spin-orbit interaction},}\ }\href {\doibase
  10.1103/PhysRevB.96.115404} {\bibfield  {journal} {\bibinfo  {journal} {Phys.
  Rev. B}\ }\textbf {\bibinfo {volume} {96}},\ \bibinfo {pages} {115404}
  (\bibinfo {year} {2017})}\BibitemShut {NoStop}%
\bibitem [{\citenamefont {Wysokinski}\ and\ \citenamefont
  {Spalek}(2013)}]{SJozef}%
  \BibitemOpen
  \bibfield  {author} {\bibinfo {author} {\bibfnamefont {Marcin~M.}\
  \bibnamefont {Wysokinski}}\ and\ \bibinfo {author} {\bibfnamefont {Jozef}\
  \bibnamefont {Spalek}},\ }\bibfield  {title} {\enquote {\bibinfo {title}
  {Seebeck effect in the graphene-superconductor junction},}\ }\href {\doibase
  10.1063/1.4802503} {\bibfield  {journal} {\bibinfo  {journal} {Journal of
  Applied Physics}\ }\textbf {\bibinfo {volume} {113}},\ \bibinfo {pages}
  {163905} (\bibinfo {year} {2013})}\BibitemShut {NoStop}%
\bibitem [{\citenamefont {Zebarjadi}\ \emph {et~al.}(2012)\citenamefont
  {Zebarjadi}, \citenamefont {Esfarjani}, \citenamefont {Dresselhaus},
  \citenamefont {Ren},\ and\ \citenamefont {Chen}}]{C1EE02497C}%
  \BibitemOpen
  \bibfield  {author} {\bibinfo {author} {\bibfnamefont {M.}~\bibnamefont
  {Zebarjadi}}, \bibinfo {author} {\bibfnamefont {K.}~\bibnamefont
  {Esfarjani}}, \bibinfo {author} {\bibfnamefont {M.~S.}\ \bibnamefont
  {Dresselhaus}}, \bibinfo {author} {\bibfnamefont {Z.~F.}\ \bibnamefont
  {Ren}}, \ and\ \bibinfo {author} {\bibfnamefont {G.}~\bibnamefont {Chen}},\
  }\bibfield  {title} {\enquote {\bibinfo {title} {Perspectives on
  thermoelectrics: from fundamentals to device applications},}\ }\href
  {\doibase 10.1039/C1EE02497C} {\bibfield  {journal} {\bibinfo  {journal}
  {Energy Environ. Sci.}\ }\textbf {\bibinfo {volume} {5}},\ \bibinfo {pages}
  {5147--5162} (\bibinfo {year} {2012})}\BibitemShut {NoStop}%
\bibitem [{\citenamefont {Snyder}\ and\ \citenamefont
  {Toberer}(2008)}]{Snyder2008}%
  \BibitemOpen
  \bibfield  {author} {\bibinfo {author} {\bibfnamefont {G.~Jeffrey}\
  \bibnamefont {Snyder}}\ and\ \bibinfo {author} {\bibfnamefont {Eric~S.}\
  \bibnamefont {Toberer}},\ }\bibfield  {title} {\enquote {\bibinfo {title}
  {Complex thermoelectric materials},}\ }\href {\doibase 10.1038/nmat2090}
  {\bibfield  {journal} {\bibinfo  {journal} {Nature Materials}\ }\textbf
  {\bibinfo {volume} {7}},\ \bibinfo {pages} {105--114} (\bibinfo {year}
  {2008})}\BibitemShut {NoStop}%
\bibitem [{\citenamefont {Giazotto}\ \emph {et~al.}(2006)\citenamefont
  {Giazotto}, \citenamefont {Heikkil\"a}, \citenamefont {Luukanen},
  \citenamefont {Savin},\ and\ \citenamefont {Pekola}}]{RevModPhys.78.217}%
  \BibitemOpen
  \bibfield  {author} {\bibinfo {author} {\bibfnamefont {Francesco}\
  \bibnamefont {Giazotto}}, \bibinfo {author} {\bibfnamefont {Tero~T.}\
  \bibnamefont {Heikkil\"a}}, \bibinfo {author} {\bibfnamefont {Arttu}\
  \bibnamefont {Luukanen}}, \bibinfo {author} {\bibfnamefont {Alexander~M.}\
  \bibnamefont {Savin}}, \ and\ \bibinfo {author} {\bibfnamefont {Jukka~P.}\
  \bibnamefont {Pekola}},\ }\bibfield  {title} {\enquote {\bibinfo {title}
  {Opportunities for mesoscopics in thermometry and refrigeration: Physics and
  applications},}\ }\href {\doibase 10.1103/RevModPhys.78.217} {\bibfield
  {journal} {\bibinfo  {journal} {Rev. Mod. Phys.}\ }\textbf {\bibinfo {volume}
  {78}},\ \bibinfo {pages} {217--274} (\bibinfo {year} {2006})}\BibitemShut
  {NoStop}%
\bibitem [{\citenamefont {Yokoyama}\ \emph {et~al.}(2008)\citenamefont
  {Yokoyama}, \citenamefont {Linder},\ and\ \citenamefont
  {Sudb\o{}}}]{Linder2008}%
  \BibitemOpen
  \bibfield  {author} {\bibinfo {author} {\bibfnamefont {Takehito}\
  \bibnamefont {Yokoyama}}, \bibinfo {author} {\bibfnamefont {Jacob}\
  \bibnamefont {Linder}}, \ and\ \bibinfo {author} {\bibfnamefont {Asle}\
  \bibnamefont {Sudb\o{}}},\ }\bibfield  {title} {\enquote {\bibinfo {title}
  {Heat transport by dirac fermions in normal/superconducting graphene
  junctions},}\ }\href {\doibase 10.1103/PhysRevB.77.132503} {\bibfield
  {journal} {\bibinfo  {journal} {Phys. Rev. B}\ }\textbf {\bibinfo {volume}
  {77}},\ \bibinfo {pages} {132503} (\bibinfo {year} {2008})}\BibitemShut
  {NoStop}%
\bibitem [{\citenamefont {Paul}\ \emph {et~al.}(2016)\citenamefont {Paul},
  \citenamefont {Sarkar},\ and\ \citenamefont {Saha}}]{Arijit2016}%
  \BibitemOpen
  \bibfield  {author} {\bibinfo {author} {\bibfnamefont {Ganesh~C.}\
  \bibnamefont {Paul}}, \bibinfo {author} {\bibfnamefont {Surajit}\
  \bibnamefont {Sarkar}}, \ and\ \bibinfo {author} {\bibfnamefont {Arijit}\
  \bibnamefont {Saha}},\ }\bibfield  {title} {\enquote {\bibinfo {title}
  {Thermal conductance by dirac fermions in a normal-insulator-superconductor
  junction of silicene},}\ }\href {\doibase 10.1103/PhysRevB.94.155453}
  {\bibfield  {journal} {\bibinfo  {journal} {Phys. Rev. B}\ }\textbf {\bibinfo
  {volume} {94}},\ \bibinfo {pages} {155453} (\bibinfo {year}
  {2016})}\BibitemShut {NoStop}%
\bibitem [{\citenamefont {Beiranvand}\ and\ \citenamefont
  {Hamzehpour}(2017)}]{Razieh}%
  \BibitemOpen
  \bibfield  {author} {\bibinfo {author} {\bibfnamefont {Razieh}\ \bibnamefont
  {Beiranvand}}\ and\ \bibinfo {author} {\bibfnamefont {Hossein}\ \bibnamefont
  {Hamzehpour}},\ }\bibfield  {title} {\enquote {\bibinfo {title}
  {Spin-dependent thermoelectric effects in graphene-based superconductor
  junctions},}\ }\href {\doibase 10.1063/1.4976005} {\bibfield  {journal}
  {\bibinfo  {journal} {Journal of Applied Physics}\ }\textbf {\bibinfo
  {volume} {121}},\ \bibinfo {pages} {063903} (\bibinfo {year}
  {2017})}\BibitemShut {NoStop}%
\bibitem [{\citenamefont {Kim}\ \emph {et~al.}(2016)\citenamefont {Kim},
  \citenamefont {Park},\ and\ \citenamefont {Marzari}}]{Marzari2016}%
  \BibitemOpen
  \bibfield  {author} {\bibinfo {author} {\bibfnamefont {T.~Y.}\ \bibnamefont
  {Kim}}, \bibinfo {author} {\bibfnamefont {C.-H.}\ \bibnamefont {Park}}, \
  and\ \bibinfo {author} {\bibfnamefont {N.}~\bibnamefont {Marzari}},\
  }\bibfield  {title} {\enquote {\bibinfo {title} {The electronic thermal
  conductivity of graphene},}\ }\href
  {https://doi.org/10.1021/acs.nanolett.5b05288} {\bibfield  {journal}
  {\bibinfo  {journal} {Nano Lett.}\ }\textbf {\bibinfo {volume} {16}},\
  \bibinfo {pages} {2439--2443} (\bibinfo {year} {2016})}\BibitemShut {NoStop}%
\bibitem [{\citenamefont {Sevin\ifmmode~\mbox{\c{c}}\else \c{c}\fi{}li}\ and\
  \citenamefont {Cuniberti}(2010)}]{PhysRevB.81.113401}%
  \BibitemOpen
  \bibfield  {author} {\bibinfo {author} {\bibfnamefont {H.}~\bibnamefont
  {Sevin\ifmmode~\mbox{\c{c}}\else \c{c}\fi{}li}}\ and\ \bibinfo {author}
  {\bibfnamefont {G.}~\bibnamefont {Cuniberti}},\ }\bibfield  {title} {\enquote
  {\bibinfo {title} {Enhanced thermoelectric figure of merit in edge-disordered
  zigzag graphene nanoribbons},}\ }\href {\doibase 10.1103/PhysRevB.81.113401}
  {\bibfield  {journal} {\bibinfo  {journal} {Phys. Rev. B}\ }\textbf {\bibinfo
  {volume} {81}},\ \bibinfo {pages} {113401} (\bibinfo {year}
  {2010})}\BibitemShut {NoStop}%
\bibitem [{\citenamefont {Zare}(2019)}]{Zare2019}%
  \BibitemOpen
  \bibfield  {author} {\bibinfo {author} {\bibfnamefont {Moslem}\ \bibnamefont
  {Zare}},\ }\bibfield  {title} {\enquote {\bibinfo {title} {Negative
  differential thermal conductance in a borophane normal
  metal{\textendash}superconductor junction},}\ }\href {\doibase
  10.1088/1361-6668/ab3caf} {\bibfield  {journal} {\bibinfo  {journal}
  {Superconductor Science and Technology}\ }\textbf {\bibinfo {volume} {32}},\
  \bibinfo {pages} {115002} (\bibinfo {year} {2019})}\BibitemShut {NoStop}%
\bibitem [{\citenamefont {Uchida}\ \emph {et~al.}(2014)\citenamefont {Uchida},
  \citenamefont {Habe},\ and\ \citenamefont {Asano}}]{Uchida2014}%
  \BibitemOpen
  \bibfield  {author} {\bibinfo {author} {\bibfnamefont {Shuhei}\ \bibnamefont
  {Uchida}}, \bibinfo {author} {\bibfnamefont {Tetsuro}\ \bibnamefont {Habe}},
  \ and\ \bibinfo {author} {\bibfnamefont {Yasuhiro}\ \bibnamefont {Asano}},\
  }\bibfield  {title} {\enquote {\bibinfo {title} {Andreev reflection in weyl
  semimetals},}\ }\href {\doibase 10.7566/JPSJ.83.064711} {\bibfield  {journal}
  {\bibinfo  {journal} {Journal of the Physical Society of Japan}\ }\textbf
  {\bibinfo {volume} {83}},\ \bibinfo {pages} {064711} (\bibinfo {year}
  {2014})}\BibitemShut {NoStop}%
\bibitem [{\citenamefont {Khanna}\ \emph {et~al.}(2016)\citenamefont {Khanna},
  \citenamefont {Mukherjee}, \citenamefont {Kundu},\ and\ \citenamefont
  {Rao}}]{SRAK2016}%
  \BibitemOpen
  \bibfield  {author} {\bibinfo {author} {\bibfnamefont {Udit}\ \bibnamefont
  {Khanna}}, \bibinfo {author} {\bibfnamefont {Dibya~Kanti}\ \bibnamefont
  {Mukherjee}}, \bibinfo {author} {\bibfnamefont {Arijit}\ \bibnamefont
  {Kundu}}, \ and\ \bibinfo {author} {\bibfnamefont {Sumathi}\ \bibnamefont
  {Rao}},\ }\bibfield  {title} {\enquote {\bibinfo {title} {Chiral nodes and
  oscillations in the josephson current in weyl semimetals},}\ }\href {\doibase
  10.1103/PhysRevB.93.121409} {\bibfield  {journal} {\bibinfo  {journal} {Phys.
  Rev. B}\ }\textbf {\bibinfo {volume} {93}},\ \bibinfo {pages} {121409}
  (\bibinfo {year} {2016})}\BibitemShut {NoStop}%
\bibitem [{\citenamefont {Mukherjee}\ \emph {et~al.}(2017)\citenamefont
  {Mukherjee}, \citenamefont {Rao},\ and\ \citenamefont {Kundu}}]{SRAK2017}%
  \BibitemOpen
  \bibfield  {author} {\bibinfo {author} {\bibfnamefont {Dibya~Kanti}\
  \bibnamefont {Mukherjee}}, \bibinfo {author} {\bibfnamefont {Sumathi}\
  \bibnamefont {Rao}}, \ and\ \bibinfo {author} {\bibfnamefont {Arijit}\
  \bibnamefont {Kundu}},\ }\bibfield  {title} {\enquote {\bibinfo {title}
  {Transport through andreev bound states in a weyl semimetal quantum dot},}\
  }\href {\doibase 10.1103/PhysRevB.96.161408} {\bibfield  {journal} {\bibinfo
  {journal} {Phys. Rev. B}\ }\textbf {\bibinfo {volume} {96}},\ \bibinfo
  {pages} {161408} (\bibinfo {year} {2017})}\BibitemShut {NoStop}%
\bibitem [{\citenamefont {Zhang}\ \emph
  {et~al.}(2018{\natexlab{a}})\citenamefont {Zhang}, \citenamefont {Dolcini},
  \citenamefont {Breunig},\ and\ \citenamefont {Trauzettel}}]{Trauzettel2018}%
  \BibitemOpen
  \bibfield  {author} {\bibinfo {author} {\bibfnamefont {Song-Bo}\ \bibnamefont
  {Zhang}}, \bibinfo {author} {\bibfnamefont {Fabrizio}\ \bibnamefont
  {Dolcini}}, \bibinfo {author} {\bibfnamefont {Daniel}\ \bibnamefont
  {Breunig}}, \ and\ \bibinfo {author} {\bibfnamefont {Bj\"orn}\ \bibnamefont
  {Trauzettel}},\ }\bibfield  {title} {\enquote {\bibinfo {title} {Appearance
  of the universal value $e^2/h$ of the zero-bias conductance in a weyl
  semimetal-superconductor junction},}\ }\href {\doibase
  10.1103/PhysRevB.97.041116} {\bibfield  {journal} {\bibinfo  {journal} {Phys.
  Rev. B}\ }\textbf {\bibinfo {volume} {97}},\ \bibinfo {pages} {041116}
  (\bibinfo {year} {2018}{\natexlab{a}})}\BibitemShut {NoStop}%
\bibitem [{\citenamefont {Zheng}\ \emph {et~al.}(2021)\citenamefont {Zheng},
  \citenamefont {Chen},\ and\ \citenamefont {Xing}}]{PhysRevB.104.075420}%
  \BibitemOpen
  \bibfield  {author} {\bibinfo {author} {\bibfnamefont {Yue}\ \bibnamefont
  {Zheng}}, \bibinfo {author} {\bibfnamefont {Wei}\ \bibnamefont {Chen}}, \
  and\ \bibinfo {author} {\bibfnamefont {D.~Y.}\ \bibnamefont {Xing}},\
  }\bibfield  {title} {\enquote {\bibinfo {title} {Andreev reflection in
  fermi-arc surface states of weyl semimetals},}\ }\href {\doibase
  10.1103/PhysRevB.104.075420} {\bibfield  {journal} {\bibinfo  {journal}
  {Phys. Rev. B}\ }\textbf {\bibinfo {volume} {104}},\ \bibinfo {pages}
  {075420} (\bibinfo {year} {2021})}\BibitemShut {NoStop}%
\bibitem [{\citenamefont {Dutta}\ \emph {et~al.}(2020)\citenamefont {Dutta},
  \citenamefont {Parhizgar},\ and\ \citenamefont
  {Black-Schaffer}}]{Paramita2020}%
  \BibitemOpen
  \bibfield  {author} {\bibinfo {author} {\bibfnamefont {Paramita}\
  \bibnamefont {Dutta}}, \bibinfo {author} {\bibfnamefont {Fariborz}\
  \bibnamefont {Parhizgar}}, \ and\ \bibinfo {author} {\bibfnamefont
  {Annica~M.}\ \bibnamefont {Black-Schaffer}},\ }\bibfield  {title} {\enquote
  {\bibinfo {title} {Finite bulk josephson currents and chirality blockade
  removal from interorbital pairing in magnetic weyl semimetals},}\ }\href
  {\doibase 10.1103/PhysRevB.101.064514} {\bibfield  {journal} {\bibinfo
  {journal} {Phys. Rev. B}\ }\textbf {\bibinfo {volume} {101}},\ \bibinfo
  {pages} {064514} (\bibinfo {year} {2020})}\BibitemShut {NoStop}%
\bibitem [{\citenamefont {Stockert}\ \emph {et~al.}(2017)\citenamefont
  {Stockert}, \citenamefont {dos Reis}, \citenamefont {Ajeesh}, \citenamefont
  {Watzman}, \citenamefont {Schmidt}, \citenamefont {Shekhar}, \citenamefont
  {Heremans}, \citenamefont {Felser}, \citenamefont {Baenitz},\ and\
  \citenamefont {Nicklas}}]{Nicklas2017}%
  \BibitemOpen
  \bibfield  {author} {\bibinfo {author} {\bibfnamefont {U.}~\bibnamefont
  {Stockert}}, \bibinfo {author} {\bibfnamefont {R.~D.}\ \bibnamefont {dos
  Reis}}, \bibinfo {author} {\bibfnamefont {M.~O.}\ \bibnamefont {Ajeesh}},
  \bibinfo {author} {\bibfnamefont {S.~J.}\ \bibnamefont {Watzman}}, \bibinfo
  {author} {\bibfnamefont {M.}~\bibnamefont {Schmidt}}, \bibinfo {author}
  {\bibfnamefont {C.}~\bibnamefont {Shekhar}}, \bibinfo {author} {\bibfnamefont
  {J.~P.}\ \bibnamefont {Heremans}}, \bibinfo {author} {\bibfnamefont
  {C.}~\bibnamefont {Felser}}, \bibinfo {author} {\bibfnamefont
  {M.}~\bibnamefont {Baenitz}}, \ and\ \bibinfo {author} {\bibfnamefont
  {M.}~\bibnamefont {Nicklas}},\ }\bibfield  {title} {\enquote {\bibinfo
  {title} {Thermopower and thermal conductivity in the weyl semimetal {NbP}},}\
  }\href {\doibase 10.1088/1361-648x/aa7a3b} {\bibfield  {journal} {\bibinfo
  {journal} {J. Phys.: Condens. Matter}\ }\textbf {\bibinfo {volume} {29}},\
  \bibinfo {pages} {325701} (\bibinfo {year} {2017})}\BibitemShut {NoStop}%
\bibitem [{\citenamefont {Chen}\ and\ \citenamefont
  {Fiete}(2016)}]{Gregory2016}%
  \BibitemOpen
  \bibfield  {author} {\bibinfo {author} {\bibfnamefont {Qi}~\bibnamefont
  {Chen}}\ and\ \bibinfo {author} {\bibfnamefont {Gregory~A.}\ \bibnamefont
  {Fiete}},\ }\bibfield  {title} {\enquote {\bibinfo {title} {Thermoelectric
  transport in double-weyl semimetals},}\ }\href {\doibase
  10.1103/PhysRevB.93.155125} {\bibfield  {journal} {\bibinfo  {journal} {Phys.
  Rev. B}\ }\textbf {\bibinfo {volume} {93}},\ \bibinfo {pages} {155125}
  (\bibinfo {year} {2016})}\BibitemShut {NoStop}%
\bibitem [{\citenamefont {Lundgren}\ \emph {et~al.}(2014)\citenamefont
  {Lundgren}, \citenamefont {Laurell},\ and\ \citenamefont
  {Fiete}}]{Gregory2014}%
  \BibitemOpen
  \bibfield  {author} {\bibinfo {author} {\bibfnamefont {Rex}\ \bibnamefont
  {Lundgren}}, \bibinfo {author} {\bibfnamefont {Pontus}\ \bibnamefont
  {Laurell}}, \ and\ \bibinfo {author} {\bibfnamefont {Gregory~A.}\
  \bibnamefont {Fiete}},\ }\bibfield  {title} {\enquote {\bibinfo {title}
  {Thermoelectric properties of weyl and dirac semimetals},}\ }\href {\doibase
  10.1103/PhysRevB.90.165115} {\bibfield  {journal} {\bibinfo  {journal} {Phys.
  Rev. B}\ }\textbf {\bibinfo {volume} {90}},\ \bibinfo {pages} {165115}
  (\bibinfo {year} {2014})}\BibitemShut {NoStop}%
\bibitem [{\citenamefont {Meng}\ and\ \citenamefont
  {Balents}(2012)}]{PhysRevB.86.054504}%
  \BibitemOpen
  \bibfield  {author} {\bibinfo {author} {\bibfnamefont {Tobias}\ \bibnamefont
  {Meng}}\ and\ \bibinfo {author} {\bibfnamefont {Leon}\ \bibnamefont
  {Balents}},\ }\bibfield  {title} {\enquote {\bibinfo {title} {Weyl
  superconductors},}\ }\href {\doibase 10.1103/PhysRevB.86.054504} {\bibfield
  {journal} {\bibinfo  {journal} {Phys. Rev. B}\ }\textbf {\bibinfo {volume}
  {86}},\ \bibinfo {pages} {054504} (\bibinfo {year} {2012})}\BibitemShut
  {NoStop}%
\bibitem [{\citenamefont {Cho}\ \emph {et~al.}(2012)\citenamefont {Cho},
  \citenamefont {Bardarson}, \citenamefont {Lu},\ and\ \citenamefont
  {Moore}}]{PhysRevB.86.214514}%
  \BibitemOpen
  \bibfield  {author} {\bibinfo {author} {\bibfnamefont {Gil~Young}\
  \bibnamefont {Cho}}, \bibinfo {author} {\bibfnamefont {Jens~H.}\ \bibnamefont
  {Bardarson}}, \bibinfo {author} {\bibfnamefont {Yuan-Ming}\ \bibnamefont
  {Lu}}, \ and\ \bibinfo {author} {\bibfnamefont {Joel~E.}\ \bibnamefont
  {Moore}},\ }\bibfield  {title} {\enquote {\bibinfo {title} {Superconductivity
  of doped weyl semimetals: Finite-momentum pairing and electronic analog of
  the ${}^{3}$he-$a$ phase},}\ }\href {\doibase 10.1103/PhysRevB.86.214514}
  {\bibfield  {journal} {\bibinfo  {journal} {Phys. Rev. B}\ }\textbf {\bibinfo
  {volume} {86}},\ \bibinfo {pages} {214514} (\bibinfo {year}
  {2012})}\BibitemShut {NoStop}%
\bibitem [{\citenamefont {Bednik}\ \emph {et~al.}(2015)\citenamefont {Bednik},
  \citenamefont {Zyuzin},\ and\ \citenamefont {Burkov}}]{PhysRevB.92.035153}%
  \BibitemOpen
  \bibfield  {author} {\bibinfo {author} {\bibfnamefont {G.}~\bibnamefont
  {Bednik}}, \bibinfo {author} {\bibfnamefont {A.~A.}\ \bibnamefont {Zyuzin}},
  \ and\ \bibinfo {author} {\bibfnamefont {A.~A.}\ \bibnamefont {Burkov}},\
  }\bibfield  {title} {\enquote {\bibinfo {title} {Superconductivity in weyl
  metals},}\ }\href {\doibase 10.1103/PhysRevB.92.035153} {\bibfield  {journal}
  {\bibinfo  {journal} {Phys. Rev. B}\ }\textbf {\bibinfo {volume} {92}},\
  \bibinfo {pages} {035153} (\bibinfo {year} {2015})}\BibitemShut {NoStop}%
\bibitem [{\citenamefont {Bovenzi}\ \emph {et~al.}(2017)\citenamefont
  {Bovenzi}, \citenamefont {Breitkreiz}, \citenamefont {Baireuther},
  \citenamefont {O'Brien}, \citenamefont {Tworzyd\l{}o}, \citenamefont
  {Adagideli},\ and\ \citenamefont {Beenakker}}]{CBBeenakker}%
  \BibitemOpen
  \bibfield  {author} {\bibinfo {author} {\bibfnamefont {N.}~\bibnamefont
  {Bovenzi}}, \bibinfo {author} {\bibfnamefont {M.}~\bibnamefont {Breitkreiz}},
  \bibinfo {author} {\bibfnamefont {P.}~\bibnamefont {Baireuther}}, \bibinfo
  {author} {\bibfnamefont {T.~E.}\ \bibnamefont {O'Brien}}, \bibinfo {author}
  {\bibfnamefont {J.}~\bibnamefont {Tworzyd\l{}o}}, \bibinfo {author}
  {\bibfnamefont {\ifmmode \dot{I}\else~\.{I}\fi{}.}\ \bibnamefont
  {Adagideli}}, \ and\ \bibinfo {author} {\bibfnamefont {C.~W.~J.}\
  \bibnamefont {Beenakker}},\ }\bibfield  {title} {\enquote {\bibinfo {title}
  {Chirality blockade of andreev reflection in a magnetic weyl semimetal},}\
  }\href {\doibase 10.1103/PhysRevB.96.035437} {\bibfield  {journal} {\bibinfo
  {journal} {Phys. Rev. B}\ }\textbf {\bibinfo {volume} {96}},\ \bibinfo
  {pages} {035437} (\bibinfo {year} {2017})}\BibitemShut {NoStop}%
\bibitem [{\citenamefont {Zhang}\ \emph
  {et~al.}(2018{\natexlab{b}})\citenamefont {Zhang}, \citenamefont
  {Erdmenger},\ and\ \citenamefont {Trauzettel}}]{Trauzettel_PRL}%
  \BibitemOpen
  \bibfield  {author} {\bibinfo {author} {\bibfnamefont {Song-Bo}\ \bibnamefont
  {Zhang}}, \bibinfo {author} {\bibfnamefont {Johanna}\ \bibnamefont
  {Erdmenger}}, \ and\ \bibinfo {author} {\bibfnamefont {Bj\"{ö}rn}\
  \bibnamefont {Trauzettel}},\ }\bibfield  {title} {\enquote {\bibinfo {title}
  {Chirality josephson current due to a novel quantum anomaly in
  inversion-asymmetric weyl semimetals},}\ }\href {\doibase
  https://doi.org/10.1103/PhysRevLett.121.226604} {\bibfield  {journal}
  {\bibinfo  {journal} {PRL}\ }\textbf {\bibinfo {volume} {121}},\ \bibinfo
  {pages} {226604} (\bibinfo {year} {2018}{\natexlab{b}})}\BibitemShut
  {NoStop}%
\bibitem [{\citenamefont {Beenakker}(2006)}]{SARBeenakker}%
  \BibitemOpen
  \bibfield  {author} {\bibinfo {author} {\bibfnamefont {C.~W.~J.}\
  \bibnamefont {Beenakker}},\ }\bibfield  {title} {\enquote {\bibinfo {title}
  {Specular andreev reflection in graphene},}\ }\href {\doibase
  10.1103/PhysRevLett.97.067007} {\bibfield  {journal} {\bibinfo  {journal}
  {Phys. Rev. Lett.}\ }\textbf {\bibinfo {volume} {97}},\ \bibinfo {pages}
  {067007} (\bibinfo {year} {2006})}\BibitemShut {NoStop}%
\bibitem [{\citenamefont {Blonder}\ \emph {et~al.}(1982)\citenamefont
  {Blonder}, \citenamefont {Tinkham},\ and\ \citenamefont {Klapwijk}}]{BTK}%
  \BibitemOpen
  \bibfield  {author} {\bibinfo {author} {\bibfnamefont {G.~E.}\ \bibnamefont
  {Blonder}}, \bibinfo {author} {\bibfnamefont {M.}~\bibnamefont {Tinkham}}, \
  and\ \bibinfo {author} {\bibfnamefont {T.~M.}\ \bibnamefont {Klapwijk}},\
  }\bibfield  {title} {\enquote {\bibinfo {title} {Transition from metallic to
  tunneling regimes in superconducting microconstrictions: Excess current,
  charge imbalance, and supercurrent conversion},}\ }\href {\doibase
  10.1103/PhysRevB.25.4515} {\bibfield  {journal} {\bibinfo  {journal} {Phys.
  Rev. B}\ }\textbf {\bibinfo {volume} {25}},\ \bibinfo {pages} {4515--4532}
  (\bibinfo {year} {1982})}\BibitemShut {NoStop}%
\bibitem [{\citenamefont {Riedel}\ and\ \citenamefont
  {Bagwell}(1993)}]{Bagwell1993}%
  \BibitemOpen
  \bibfield  {author} {\bibinfo {author} {\bibfnamefont {Richard~A.}\
  \bibnamefont {Riedel}}\ and\ \bibinfo {author} {\bibfnamefont {Philip~F.}\
  \bibnamefont {Bagwell}},\ }\bibfield  {title} {\enquote {\bibinfo {title}
  {Current-voltage relation of a normal-metal--superconductor junction},}\
  }\href {\doibase 10.1103/PhysRevB.48.15198} {\bibfield  {journal} {\bibinfo
  {journal} {Phys. Rev. B}\ }\textbf {\bibinfo {volume} {48}},\ \bibinfo
  {pages} {15198--15208} (\bibinfo {year} {1993})}\BibitemShut {NoStop}%
\bibitem [{\citenamefont {Sivan}\ and\ \citenamefont
  {Imry}(1986)}]{Sivan_Imry1993}%
  \BibitemOpen
  \bibfield  {author} {\bibinfo {author} {\bibfnamefont {U.}~\bibnamefont
  {Sivan}}\ and\ \bibinfo {author} {\bibfnamefont {Y.}~\bibnamefont {Imry}},\
  }\bibfield  {title} {\enquote {\bibinfo {title} {Multichannel landauer
  formula for thermoelectric transport with application to thermopower near the
  mobility edge},}\ }\href {\doibase 10.1103/PhysRevB.33.551} {\bibfield
  {journal} {\bibinfo  {journal} {Phys. Rev. B}\ }\textbf {\bibinfo {volume}
  {33}},\ \bibinfo {pages} {551--558} (\bibinfo {year} {1986})}\BibitemShut
  {NoStop}%
\bibitem [{\citenamefont {Ashcroft}\ \emph {et~al.}(1976)\citenamefont
  {Ashcroft}, \citenamefont {Mermin},\ and\ \citenamefont {D.}}]{Ashcroft76}%
  \BibitemOpen
  \bibfield  {author} {\bibinfo {author} {\bibnamefont {Ashcroft}}, \bibinfo
  {author} {\bibfnamefont {N.~W.}\ \bibnamefont {Mermin}}, \ and\ \bibinfo
  {author} {\bibfnamefont {N.}~\bibnamefont {D.}},\ }\href@noop {} {\emph
  {\bibinfo {title} {Solid State Physics}}}\ (\bibinfo  {publisher}
  {Holt-Saunders},\ \bibinfo {year} {1976})\BibitemShut {NoStop}%
\bibitem [{\citenamefont {Goldsmid}(1956)}]{Goldsmid1956}%
  \BibitemOpen
  \bibfield  {author} {\bibinfo {author} {\bibfnamefont {H~J}\ \bibnamefont
  {Goldsmid}},\ }\bibfield  {title} {\enquote {\bibinfo {title} {The thermal
  conductivity of bismuth telluride},}\ }\href {\doibase
  10.1088/0370-1301/69/2/310} {\bibfield  {journal} {\bibinfo  {journal} {Proc.
  Phys. Soc. B}\ }\textbf {\bibinfo {volume} {69}},\ \bibinfo {pages}
  {203--209} (\bibinfo {year} {1956})}\BibitemShut {NoStop}%
\bibitem [{\citenamefont {Crossno}\ \emph {et~al.}(2016)\citenamefont
  {Crossno}, \citenamefont {Shi}, \citenamefont {Wang}, \citenamefont {Liu},
  \citenamefont {Harzheim}, \citenamefont {Lucas}, \citenamefont {Sachdev},
  \citenamefont {Kim}, \citenamefont {Taniguchi}, \citenamefont {Watanabe},
  \citenamefont {Ohki},\ and\ \citenamefont {Fong}}]{Crassno2016}%
  \BibitemOpen
  \bibfield  {author} {\bibinfo {author} {\bibfnamefont {J.}~\bibnamefont
  {Crossno}}, \bibinfo {author} {\bibfnamefont {J.~K.}\ \bibnamefont {Shi}},
  \bibinfo {author} {\bibfnamefont {K.}~\bibnamefont {Wang}}, \bibinfo {author}
  {\bibfnamefont {X.}~\bibnamefont {Liu}}, \bibinfo {author} {\bibfnamefont
  {A.}~\bibnamefont {Harzheim}}, \bibinfo {author} {\bibfnamefont
  {A.}~\bibnamefont {Lucas}}, \bibinfo {author} {\bibfnamefont
  {S.}~\bibnamefont {Sachdev}}, \bibinfo {author} {\bibfnamefont
  {P.}~\bibnamefont {Kim}}, \bibinfo {author} {\bibfnamefont {T.}~\bibnamefont
  {Taniguchi}}, \bibinfo {author} {\bibfnamefont {K.}~\bibnamefont {Watanabe}},
  \bibinfo {author} {\bibfnamefont {T.~A.}\ \bibnamefont {Ohki}}, \ and\
  \bibinfo {author} {\bibfnamefont {K.~C.}\ \bibnamefont {Fong}},\ }\bibfield
  {title} {\enquote {\bibinfo {title} {Observation of the dirac fluid and the
  breakdown of the wiedemann-franz law in graphene},}\ }\href {DOI:
  10.1126/science.aad0343} {\bibfield  {journal} {\bibinfo  {journal}
  {Science}\ }\textbf {\bibinfo {volume} {351}},\ \bibinfo {pages} {1058--1061}
  (\bibinfo {year} {2016})}\BibitemShut {NoStop}%
\bibitem [{\citenamefont {Wakeham}\ \emph {et~al.}(2011)\citenamefont
  {Wakeham}, \citenamefont {Bangura}, \citenamefont {Xu}, \citenamefont
  {Mercure}, \citenamefont {Greenblatt},\ and\ \citenamefont
  {Hussey}}]{Wakeham2011}%
  \BibitemOpen
  \bibfield  {author} {\bibinfo {author} {\bibfnamefont {N.}~\bibnamefont
  {Wakeham}}, \bibinfo {author} {\bibfnamefont {A.}~\bibnamefont {Bangura}},
  \bibinfo {author} {\bibfnamefont {X.}~\bibnamefont {Xu}}, \bibinfo {author}
  {\bibfnamefont {J.~F.}\ \bibnamefont {Mercure}}, \bibinfo {author}
  {\bibfnamefont {M.}~\bibnamefont {Greenblatt}}, \ and\ \bibinfo {author}
  {\bibfnamefont {N.~E.}\ \bibnamefont {Hussey}},\ }\bibfield  {title}
  {\enquote {\bibinfo {title} {Gross violation of the {W}iedemann-{F}ranz law
  in a quasi one dimensional conductor},}\ }\href
  {https://doi.org/10.1038/ncomms1406} {\bibfield  {journal} {\bibinfo
  {journal} {Nature Communications}\ }\textbf {\bibinfo {volume} {2}},\
  \bibinfo {pages} {396} (\bibinfo {year} {2011})}\BibitemShut {NoStop}%
\bibitem [{\citenamefont {Tanatar}\ \emph {et~al.}(2007)\citenamefont
  {Tanatar}, \citenamefont {Paglione}, \citenamefont {Petrovic},\ and\
  \citenamefont {Taillefer}}]{Tanatar2007}%
  \BibitemOpen
  \bibfield  {author} {\bibinfo {author} {\bibfnamefont {M.~A.}\ \bibnamefont
  {Tanatar}}, \bibinfo {author} {\bibfnamefont {J.}~\bibnamefont {Paglione}},
  \bibinfo {author} {\bibfnamefont {C.}~\bibnamefont {Petrovic}}, \ and\
  \bibinfo {author} {\bibfnamefont {L.}~\bibnamefont {Taillefer}},\ }\bibfield
  {title} {\enquote {\bibinfo {title} {Anisotropic violation of the
  {W}iedemann-{F}ranz law at a quantum critical point},}\ }\href {DOI:
  10.1126/science.1140762} {\bibfield  {journal} {\bibinfo  {journal}
  {Science}\ }\textbf {\bibinfo {volume} {316}},\ \bibinfo {pages} {1320--1322}
  (\bibinfo {year} {2007})}\BibitemShut {NoStop}%
\bibitem [{\citenamefont {Gundrum}\ \emph {et~al.}(2005)\citenamefont
  {Gundrum}, \citenamefont {Cahill},\ and\ \citenamefont
  {Averback}}]{Averback2005}%
  \BibitemOpen
  \bibfield  {author} {\bibinfo {author} {\bibfnamefont {Bryan~C.}\
  \bibnamefont {Gundrum}}, \bibinfo {author} {\bibfnamefont {David~G.}\
  \bibnamefont {Cahill}}, \ and\ \bibinfo {author} {\bibfnamefont {Robert~S.}\
  \bibnamefont {Averback}},\ }\bibfield  {title} {\enquote {\bibinfo {title}
  {Thermal conductance of metal-metal interfaces},}\ }\href {\doibase
  10.1103/PhysRevB.72.245426} {\bibfield  {journal} {\bibinfo  {journal} {Phys.
  Rev. B}\ }\textbf {\bibinfo {volume} {72}},\ \bibinfo {pages} {245426}
  (\bibinfo {year} {2005})}\BibitemShut {NoStop}%
\bibitem [{\citenamefont {Ghanbari}\ and\ \citenamefont
  {Rashedi}(2011)}]{Rashedi2011}%
  \BibitemOpen
  \bibfield  {author} {\bibinfo {author} {\bibfnamefont {R.}~\bibnamefont
  {Ghanbari}}\ and\ \bibinfo {author} {\bibfnamefont {G.}~\bibnamefont
  {Rashedi}},\ }\bibfield  {title} {\enquote {\bibinfo {title} {The
  {W}iedemann-{F}ranz law in a normal metal-superconductor junction},}\ }\href
  {\doibase 10.1088/1674-1056/20/12/127401} {\bibfield  {journal} {\bibinfo
  {journal} {Chin. Phys. B}\ }\textbf {\bibinfo {volume} {20}},\ \bibinfo
  {pages} {127401} (\bibinfo {year} {2011})}\BibitemShut {NoStop}%
\bibitem [{\citenamefont {Hesdeo}\ \emph {et~al.}(2019)\citenamefont {Hesdeo},
  \citenamefont {Krisna}, \citenamefont {Hanna}, \citenamefont {Gunara},
  \citenamefont {Hung},\ and\ \citenamefont {Nugraha}}]{Nugraha2019}%
  \BibitemOpen
  \bibfield  {author} {\bibinfo {author} {\bibfnamefont {E.~H.}\ \bibnamefont
  {Hesdeo}}, \bibinfo {author} {\bibfnamefont {L.~P.~A.}\ \bibnamefont
  {Krisna}}, \bibinfo {author} {\bibfnamefont {M.~Y.}\ \bibnamefont {Hanna}},
  \bibinfo {author} {\bibfnamefont {B.~E.}\ \bibnamefont {Gunara}}, \bibinfo
  {author} {\bibfnamefont {N.~T.}\ \bibnamefont {Hung}}, \ and\ \bibinfo
  {author} {\bibfnamefont {A.~R.~T.}\ \bibnamefont {Nugraha}},\ }\bibfield
  {title} {\enquote {\bibinfo {title} {Optimal band gap for improved
  thermoelectric performance of two-dimensional materials},}\ }\href
  {https://doi.org/10.1063/1.5100985} {\bibfield  {journal} {\bibinfo
  {journal} {J. Appl. Phys.}\ }\textbf {\bibinfo {volume} {126}},\ \bibinfo
  {pages} {035109} (\bibinfo {year} {2019})}\BibitemShut {NoStop}%
\bibitem [{\citenamefont {Rycerz}(2021)}]{AdamWFLAW2021}%
  \BibitemOpen
  \bibfield  {author} {\bibinfo {author} {\bibfnamefont {A.}~\bibnamefont
  {Rycerz}},\ }\bibfield  {title} {\enquote {\bibinfo {title}
  {Wiedemann–{F}ranz law for massless {D}irac fermions with implications for
  graphene},}\ }\href {\doibase 10.3390/ma14112704} {\bibfield  {journal}
  {\bibinfo  {journal} {Materials}\ }\textbf {\bibinfo {volume} {14}},\
  \bibinfo {pages} {2704} (\bibinfo {year} {2021})}\BibitemShut {NoStop}%
\end{thebibliography}%

\end{document}